\documentclass{aa}  

\usepackage{graphicx}
\usepackage{txfonts}
\usepackage{lipsum}
\usepackage{subcaption}         
\usepackage{lscape}             
\usepackage{placeins}           

\usepackage{xcolor}

\usepackage{hyperref}
\hypersetup{colorlinks=true,urlcolor=blue, citecolor=blue, linkcolor=blue}

\usepackage{keyval}
\usepackage{multirow}
\usepackage{todonotes}
\usepackage{soul}
\usepackage{comment}
\usepackage{orcidlink}
\usepackage{subcaption,booktabs}
\usepackage{placeins}
\usepackage{multirow}
\usepackage{natbib}
\usepackage[normalem]{ulem}

\newcommand{\Ms}{{\rm M}_\odot}

\newcommand{\sevn}{\textsc{sevn}}
\newcommand{\starfall}{\texttt{STARfall}}

\newcommand{\dragon}{\textsc{Dragon-II}}
\newcommand{\bpop}{\texttt{B-POP}}

\begin{document}

   \title{The STARfall pathway: forging intermediate-mass black holes through stellar collisions in dense star clusters}


%

   \author{Filippo Simonato\inst{1}\fnmsep\inst{2}\fnmsep\inst{3}\fnmsep\thanks{e-mail: \href{mailto:filippo.simonato@gssi.it}{filippo.simonato@gssi.it}}\orcidlink{0009-0008-5500-4160}
        \and Manuel Arca Sedda \inst{1}\fnmsep\inst{2}\fnmsep\inst{4}\fnmsep\thanks{e-mail: \href{mailto:manuel.arcasedda@gssi.it}{manuel.arcasedda@gssi.it}}
        \and Taeho Ryu\inst{5}\fnmsep\inst{6}
        \and Jakub Klencki\inst{7}
        \and Giuliano Iorio\inst{8}
        \and Sara Rastello\inst{8}\fnmsep\inst{9}
        \and Lavinia Paiella\inst{1}\fnmsep\inst{2}\fnmsep\inst{4}\orcidlink{0009-0001-7605-991X}
        \and Cristiano Ugolini\inst{1}\fnmsep\inst{2}\fnmsep\inst{4}
        \and Benedetta Mestichelli\inst{1}\fnmsep\inst{2}\orcidlink{0009-0002-1705-4729}
        \and Marica Branchesi\inst{1}\fnmsep\inst{2}\fnmsep\inst{3}
        \and Abbas Askar\inst{10}
        \and Peter Berczik\inst{10}\fnmsep\inst{11}\orcidlink{0000-0003-4176-152X}
        \and Guglielmo Costa\inst{12}\fnmsep\inst{13}
        \and Mirek Giersz\inst{10}
        \and Albrecht W. H. Kamlah\inst{14}\fnmsep\inst{15}
        \and Rainer Spurzem\inst{14}\fnmsep\inst{16}\fnmsep\inst{17}
        \and Stefano Torniamenti\inst{15}
        \and Marcelo C. Vergara\inst{18}\fnmsep\inst{19}
        }

   \institute{Gran Sasso Science Institute (GSSI), Viale Francesco Crispi 7, 67100 L’Aquila, Italy
   \and INFN - Laboratori Nazionali del Gran Sasso, 67100 Assergi, Italy
   \and INAF - Osservatorio Astronomico d’Abruzzo, Via Maggini, 64100 Teramo, Italy
   \and INAF - Osservatorio Astronomico di Roma, 00040 Monte Porzio Catone (Rome), Italy
   \and JILA, University of Colorado and National Institute of Standards and Technology, 440 UCB, Boulder, CO 80309-0440, USA
   \and Department of Astrophysical and Planetary Sciences, 391 UCB, University of Colorado, Boulder, CO 80309-0391, USA
   \and Max Planck Institute for Astrophysics, Karl-Schwarzschild-Strasse 1, 85748 Garching, Germany
   \and Institut de Ciències del Cosmos (ICCUB), Universitat de Barcelona (UB), c. Martí i Franquès, 1, 08028 Barcelona, Spain
   \and Departament de Física Quàntica i Astrofísica (FQA), Universitat de Barcelona (UB), c. Martí i Franquès 1, 08028 Barcelona, Spain
   \and Nicolaus Copernicus Astronomical Centre Polish Academy of Sciences, ul. Bartycka 18, 00-716 Warsaw, Poland
   \and Main Astronomical Observatory, National Academy of Sciences of Ukraine, 27 Akademika Zabolotnoho St, 03143 Kyiv, Ukraine
   \and Physics and Astronomy Department Galileo Galilei, University of Padova, Vicolo dell’Osservatorio 3, I–35122, Padova, Italy
   \and INAF-Osservatorio Astronomico di Padova, Vicolo dell’Osservatorio 5, Padova, Italy
   \and Astronomisches Rechen-Inst., Zentrum f\"ur Astronomie, Univ. of Heidelberg, M\"onchhofstr. 12-14, 69120, Heidelberg, Germany
   \and Max-Planck-Institut f\"ur Astronomie, K\"onigstuhl 17, 69117 Heidelberg, Germany
   \and National Astronomical Observatories and Key Laboratory of Computational Astrophysics, Chinese Academy of Sciences, 20A Datun Rd., Chaoyang District, 100101 Beijing, China
   \and Kavli Institute for Astronomy and Astrophysics, Peking University, Yiheyuan Lu 5, Haidian Qu, 100871 Beijing, China
   \and Departamento de Astronomía, Universidad de Chile, Santiago, Chile
   \and Departamento de Astronomía, Facultad de Ciencias Físicas y Matemáticas, Universidad de Concepción, Concepción, Chile}

   \date{Received September 30, 20XX}

\authorrunning{F. Simonato et al.}
 
  \abstract
    {Intermediate-mass black holes (IMBHs, $10^2 < m_{\rm BH}/\text{M}_{\odot} < 10^5$) are elusive objects falling between stellar-mass and supermassive black holes, whose origin is still highly uncertain. These objects may possibly form from the collapse of very massive stars (VMSs, $m_*\geq 150\,\Ms$) assembled through stellar collisions in compact and fast evolving star clusters.}
   {We investigate the formation of VMSs and IMBH seeds via repeated mergers and collisions as a function of cluster initial conditions, assessing the implications for globular clusters and dense high-redshift systems.}
   {We develop \starfall, a semi-analytic tool that follows the evolution of a massive stellar binary via binary-single interactions, mergers, collisions, and stellar evolution, while consistently accounting for the host cluster evolution.}
   {Our results suggest that clusters with central densities $> 10^6\,\Ms\,\text{pc}^{-3}$ undergo a catastrophic series of stellar collisions, or "starastrophe", on timescales $< 3$ Myr. Through this process, VMSs rapidly grow in mass and eventually collapse into IMBHs with masses of up to $10^3$–$10^4\,\Ms$, despite the effects of intense stellar winds. Applying our results to the compact stellar clusters observed in the Cosmic Gems Arc, we find that these environments could promote the formation of IMBHs within this mass range. Similarly, our models suggest that some Milky Way globular clusters may have hosted IMBH seeds with comparable natal masses. Finally, applying our framework to Little Red Dots, we find that, despite their longer core-collapse timescales, their inferred core densities can still support the formation of IMBHs in the same mass interval. 
   }
   {Our models show that VMS assembly requires two essential ingredients: short core-collapse times and large central densities. In the explored density range, $> 10^6\,\Ms\,\text{pc}^{-3}$, VMSs attain masses up to $10^3-10^5\,\Ms$. Compared to observed systems, our model suggest that clusters in the Cosmic Gems Arcs and several Galactic globular clusters could have nurtured VMSs that directly collapsed to IMBHs with masses $10^3-10^4\,\Ms$. 
   }


   \maketitle


\section{Stellar collisions as a pathway to intermediate-mass black holes}

Intermediate-mass black holes (IMBHs, $10^2 \leq m_{\rm BH}/\Ms < 10^5$) occupy the mass range between stellar-mass black holes (BHs) and supermassive black holes (SMBHs), yet they remain the least observationally constrained population \citep{Mezcua2017,Greene2020}. A few debated candidates have been reported in Galactic and extragalactic star clusters \citep[see e.g.,][]{Noyola2008,Bash2008,VanDerMarel2010,Lutzgendorf2013,Zocchi2016,Zocchi2019,Kiziltan2017,Askar2017,Abbate2019,Tiengo2022,Pechetti2022,DellaCroce2024}. More solid evidences include the detection of an IMBH with mass $\gtrsim (6-8) \times 10^3\,\Ms$ in $\omega$ Centauri \citep{Haberle2024,BanaresHernandez2025}, as well as gravitational-wave (GW) observations such as GW231123, which involved a merger of two black holes with a total mass of $190-265\,\Ms$ \citep{LIGO2025c}. In the coming years, next-generation GW observatories (e.g., Einstein Telescope \citealp{Branchesi2023,Abac2026}; LISA \citealp{Amaro-Seoane2017,Amaro2023}; DECIGO \citealp{Kawamura2021}) will be able to observe signals from IMBH binaries and intermediate mass-ratio inspirals, i.e., an IMBH paired with a stellar mass compact object \citep{Amaro-Seoane2007,ArcaSedda2020b,ArcaSedda2021b,Colpi2024}. Although their typical mass range remains largely unexplored, the next decades are expected to play a crucial role in unveiling the nature and origin of this elusive population.

From a theoretical point of view, explaining the formation of IMBHs through standard single-star evolution is challenging. In particular, pair-instability supernovae (PISNe) are expected to completely disrupt massive stars with a helium core mass between $64-135\,\Ms$, opening a gap in the BH mass spectrum between $\sim 100$ – $130\,\Ms$ \citep{Heger2002,Belczynski2016,Woosley2017,Spera2017,Costa2025}. Bridging this mass gap and producing more massive BHs therefore require additional formation channels (see, e.g., \citealt{Farmer2019,Farmer2020,DiCarlo2020,Costa2021,Costa2022,Ballone2023}). To consider IMBHs a distinct population rather than the extreme tail of stellar-mass BHs, processes that can sustain their formation and growth across their full mass range are required. These mechanisms are historically divided into two main complementary scenarios: high-redshift baby-galactic nuclei and dense stellar clusters.

In the ``baby-galactic nuclei'' scenario, IMBHs can form at high redshift ($z>10$) in the centers of proto-galaxies through collisions and mergers of Population III stars or their remnants \citep{Rees1984,Costa2023,Tanikawa2024}, direct collapse of massive gas clouds \citep{Silk1998,Bromm2003,Latif2013}, or a combination of both \citep{Tagawa2020}. Population III remnants are expected to produce seeds of $m_{\rm BH}\sim10^2-10^3\,\Ms$, while direct-collapse channels can yield more massive seeds, $m_{\rm BH}\sim10^4-10^6\,\Ms$. It is possible that mergers of Population III stars in star clusters produce IMBHs up to $10^4 \,\Ms$ (e.g., \citealp{Vergara2021,Wang2022,Mestichelli2024}). These objects may subsequently grow by accretion/mergers into SMBHs by $z \sim 6-8$ \citep{Shapiro2005,Volonteri2021,Volonteri2023,Maiolino2024}.

In dense stellar clusters, IMBH formation may be driven by stellar collisions and mergers \citep{PortegiesZwart2002,PortegiesZwart2004,Giersz2015,Mapelli2016,Antonini2016,ArcaSedda2020,ArcaSedda2021a,Rizzuto2021,Rizzuto2022,ArcaSedda2023,Kritos2023,GonzalezPrieto2024,Rantala2025,Rantala2026,Paiella2026}. Repeated collisions among massive stars can produce a very massive star (VMS, $m_* > 150\,\Ms$), which may collapse into an IMBH. Further growth can occur via accretion from stellar companions or mergers with stellar-mass BHs (see, e.g., \citealt{ArcaSedda2023}). Although the efficiency of IMBH seeding via VMSs depends on cluster properties such as mass segregation and core-collapse timescales \citep{Gurkan2004,FreitagII,FreitagI,Mestichelli2026}, runaway stellar collisions may enable the formation of IMBHs well above the PISN mass gap. This formation channel has been recently invoked as a possible route to form massive BH seeds in high-redshift compact clusters \citep{Guia2024,Pacucci2025,Bocchi2026,Williams2026,2026A&A...710A.235S,Liempi2026}.

Both $N$-body and Monte Carlo simulations indicate that stellar mergers and collisional processes in dense clusters can lead to the formation of IMBHs with masses up to $\sim 500\,\Ms$ \citep{Mapelli2016,Rodriguez2019,DiCarlo2019,Kremer2020,Gonzalez2021,Gonzalez2022,GonzalezPrieto2024,ArcaSedda2023a,ArcaSedda2023}. Semi-analytic studies further suggest that repeated BH mergers may produce even more massive IMBHs, reaching $10^4-10^5\,\Ms$ in dense nuclear star clusters \citep{Antonini2019,Mapelli2021,Kritos2023,Chattopadhyay2023,ArcaSedda2023}. However, many of these models typically do not follow the full dynamical evolution of the host cluster and may simplify or neglect processes such as stellar collisions or additional growth channels, with only a few notable exceptions (e.g. \citealp{Antonini2019,ArcaSedda2023,Kritos2023,Torniamenti2024,Vergara2023,Vergara2025,Rantala2026}). Interestingly, recent studies have shown that mass loss driven by optically-thick stellar winds \citep{Simonato2025,Torniamenti2026}, as well as collisions and mergers \citep{RamirezGaleano2025,Rantala2026,RomanGarza2026}, can significantly influence the final mass, and consequently the fate, of VMSs. 

Direct $N$-body simulations remain the most accurate approaches for modeling the evolution of dense stellar systems. Their main limitation, however, lies in the computational cost required to follow the full dynamical evolution, which restricts the number of realizations and typically yields only one collision chain per cluster. Indeed, the growth of massive stars is usually dominated by a single runaway object \citep{PortegiesZwart2002}, a behavior naturally expected from mass segregation: if several stars begin to grow, they rapidly sink toward the cluster core, where repeated interactions favor the emergence of one VMS.  As a result, exploring the parameter space in a statistically meaningful way remains challenging.

To overcome this limitation, we developed \starfall, a semi-analytic tool, designed to study the formation of VMSs and IMBH seeds in dense environments. \starfall\ can follow the dynamical evolution of a massive binary or of a single massive star at the cluster centre, where interactions are most frequent, while incorporating state‑of‑the‑art prescriptions for cluster evolution and dynamical encounters. Its key strength is the computational efficiency: each binary can be evolved in seconds to minutes, depending on the initial density. This enables the exploration of a much broader parameter space, including densities currently inaccessible to direct simulations.

In this work, we investigate the conditions under which runaway collisions can lead to formation of VMSs and IMBH seeds, and the implications for high-redshift dense environments, such as Little Red Dots (LRDs; \citealt{Labbe2023,Kocevski2023,Matthee2024}) and clusters in the Cosmic Gems Arc (CGA; \citealt{Adamo2024,Bradley2025,Messa2026}), as well as for present-day globular clusters (GCs; \citealt{Lutzgendorf2013,Harris2015,Kiziltan2017}), using \starfall.

In the following section, we describe the implementation of \starfall. Our results are presented in Sect. \ref{sec:results}. In Sect. \ref{sec:discussion}, we discuss their implications for VMS growth and IMBH seed formation in the CGA, GCs and LRDs. Finally, we summarize our work in Sect. \ref{sec:summary}.


\section{Dynamical evolution of massive stellar binaries with \starfall}
\label{sec:methods}

\begin{figure*} 
    \centering
    \includegraphics[width=\textwidth]{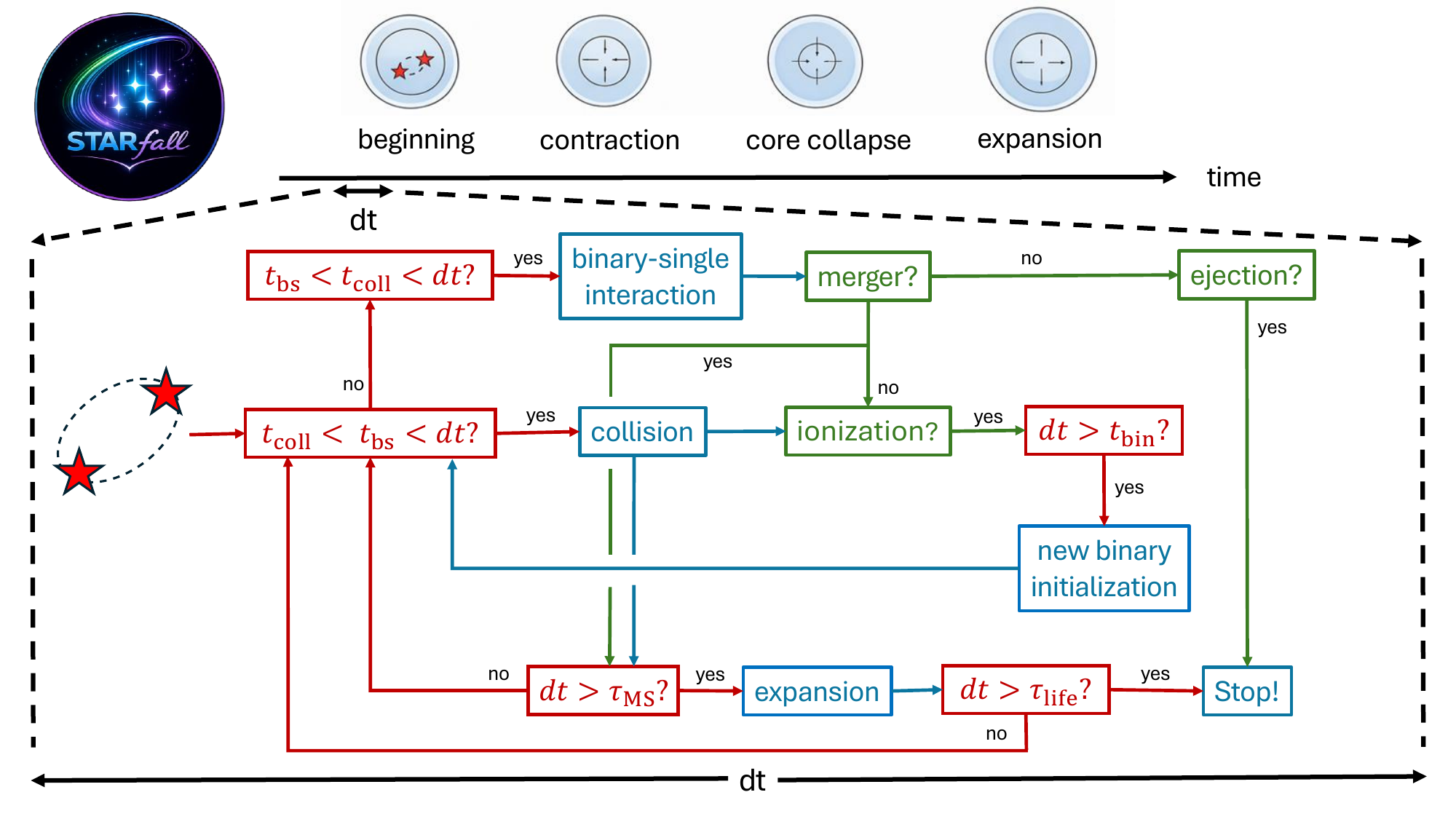}
    \caption{\starfall\ flowchart. The simulation time is advanced in logarithmic time-steps, $dt$. At each step, the collision timescale ($t_{\rm coll}$) and the binary-single interaction timescale ($t_{\rm bs}$) are computed as described in Sect. \ref{sec:dynamics}. If $t_{\rm coll} < t_{\rm bs}$, a collision occurs; otherwise, the system undergoes a binary-single interaction. If no interaction occurs during the current timestep, we proceed to the next one. If the binary is ionized after an interaction, the more massive star can experience repeated collisions or form a new binary over a timescale $t_{\rm bin} \equiv t_{\rm 3bb}$ (Eq. \ref{eq:t3bb}). After a time $\tau_{\rm MS}$ (Eq. \ref{eq:t_evo}), the star expands. The simulation terminates when the primary star collapses, i.e., when $t \geq \tau_{\rm life}$ (Eq. \ref{eq:tlife}).}
    \label{fig:flowchart}
\end{figure*}

We simulated the dynamical evolution of massive binaries, which undergo repeated binary–single interactions and stellar collisions while orbiting in the host cluster core, with the new semi-analytical tool \starfall. While \starfall\ can model the dynamical evolution of a single star, initial conditions assume a massive binary because $N$-body simulations show that primordial binaries are typically the systems that trigger the first collision chains \citep{DragonII,Rantala2025b,Mestichelli2026}. Even in the absence of primordial binaries, dense clusters form a massive binary in the core on very short timescales ($0.1-0.3$ Myr; \citealp{Rantala2024}), providing the same effective starting point for the runaway sequence. The flowchart of \starfall\ is shown in Fig. \ref{fig:flowchart}. In the following, we describe the adopted star cluster model (Sect. \ref{sec:cluster_model}), the treatment of single-star evolution (Sect. \ref{sec:stellar_evolution}), and the dynamical processes (Sect. \ref{sec:dynamics}) included in the code.

\subsection{Star cluster evolution}
\label{sec:cluster_model}
The evolution of the star cluster is modeled using the prescriptions implemented in the \bpop\footnote{\bpop\ is publicly available at this  \href{https://github.com/marcasedda/BPOP}{{github repository}}.} code \citep{ArcaSedda2026}, which tracks the time evolution of the cluster’s bound and core mass, as well as its half-mass and core radii. The main equations are reported in Appendix \ref{app:bpop}, while additional details can be found in \cite{ArcaSedda2020,ArcaSedda2023,ArcaSedda2026}. In the following, we briefly summarize the treatment of core collapse, a key physical process for our analysis. Indeed, during this phase, the steep rise in central density can trigger runaway stellar collisions that lead to the formation of VMSs \citep{PortegiesZwart2002,FreitagII,FreitagI,Gaburov2008}. 

In a semi-analytic model, the main uncertainty is related to how deep the core collapses before  interactions become energetic enough to revert the contraction. In \starfall, core collapse is halted once the core radius reaches a minimum value, $R_{\rm cc}$.
If the central massive binary remains bound, the minimum core radius is set by the binary scale:
\begin{equation}
    R_{\rm cc} = a \,\left(\displaystyle\frac{M_{\rm core}}{m_1+m_2}\right),
\end{equation}
where $a$ is the binary semi-major axis, $m_1$ and $m_2$ are the masses of the two components, and $M_{\rm core}$ is the mass of the core. This choice reflects the fact that, below this scale, the binary dominates the central dynamics and efficiently heats the surrounding cluster \citep{Binney2008}. If the binary is instead ionized, we adopted:
\begin{equation}
R_{\rm cc} = R_{\rm core}\, \left(\frac{m_1}{M_{\rm core}}\right)^{1/3},
\end{equation}
where $R_{\rm core}$ is the cluster core radius. We assume that the collapse is halted if the mass is such that the influence radius,  i.e., the region within which the gravitational potential of a single star dominates over that of the cluster, becomes larger than the wander radius, i.e., the characteristic amplitude of the star’s Brownian motion around the minimum of the cluster potential \citep{DiCintio2023,ArcaSedda2026}.

In this work, we considered star clusters with masses $\log(M_{\rm cl,0}/\Ms) = 5,6,7,$ and $8$, and half-mass radii $R_{\rm hm,0} = 0.01,\,0.05,\,0.1,\,0.5,\,1,$ and $3\ \mathrm{pc}$, with initial core densities ranging between $\rho_{\rm core,0} \sim 10^{5.5}-10^{14} \,\text{M}_{\odot}\, \text{pc}^{-3}$. These configurations allow us both to compare our results with initial conditions consistent with GCs and systems in the CGA, and to explore regions of parameter space that remain inaccessible to direct $N$-body calculations. For each combination of $M_{\rm cl,0}$ and $R_{\rm hm,0}$, we modeled 500 binaries, as described below.

\subsection{Single-star evolution}
\label{sec:stellar_evolution}
Stellar physical sizes and post–main-sequence evolution are modeled using the following prescriptions. The fitting formulae presented below are calibrated against \textsc{Parsec} stellar tracks \citep{Costa2025}, interpolated with \sevn\footnote{\sevn\ is publicly available at this\href{https://gitlab.com/sevncodes/sevn}{{ gitlab repository}}. In this work, we used version 2.15. The \textsc{Parsec} tracks are available at this \href{https://stev.oapd.inaf.it/PARSEC/tracks_v2_VMS.html}{link}.} (\texttt{SEVNtracks\_parsec\_ov05\_AGB}; \citealp{Spera2019,Iorio2023}) up to $m_{*} = 300\,\Ms$. Beyond this value, the following formulae are extrapolated. We considered metallicities $Z = 0.0002$ and $Z = 0.002$.

We first initialized stellar radii at zero-age main sequence (ZAMS) using the following mass–radius relation:
\begin{equation}
    \frac{R_{\rm ZAMS}}{1\,{\rm R{_\odot}}} = a_* \left(\frac{m_*}{1\,{\rm M}_\odot}\right)^{b_*},
    \label{eq:r_zams}
\end{equation}
and assumed it to be constant until the end of the main sequence.

To integrate stellar expansion within \starfall, we assumed that after a time:
\begin{equation}
    \frac{\tau_{\rm MS}}{1\,{\rm Myr}} = a_{\rm MS} \left(\frac{m_*}{1\,\Ms}\right)^{b_{\rm MS}} + c_{\rm MS}\,, 
    \label{eq:t_evo}
\end{equation}
the stellar radius increases by a factor $f_{\rm rad} = R_*(\tau_{\rm MS})/R_{\rm ZAMS}$, obtained from \sevn-\textsc{Parsec} tracks (see lower panel of Fig. \ref{fig:rem_mass_f_rad}). We note that the timescale over which the massive star rapidly expands is smaller than typical dynamical timescales of the cluster. The resulting radius is compared against the prescription reported in \cite{Vergara2025}:
\begin{equation}
    \frac{R_*}{1\,{\rm R_{\odot}}} = \min \left(6\times10^4+2.8\frac{m_*}{1\,\Ms}\,,\, 2\times10^5\right)\,.
\end{equation}
The final adopted stellar radius is chosen as the minimum of the two. In almost all cases, the radii reported by \cite{Vergara2025} are larger than those from \sevn-\textsc{Parsec} tracks.

We assumed that all stars have a lifetime:
\begin{equation}
    \frac{\tau_{\rm life}}{1\,{\rm Myr}} = a_{\rm life} \left(\frac{m_*}{1\,\Ms}\right)^{b_{\rm life}} + c_{\rm life}\,.
    \label{eq:tlife}
\end{equation}
By construction, the condition $\tau_{\rm MS} / \tau_{\rm life} < 1$ is always satisfied. The values of the fitting parameters are listed in Table \ref{tab1}.

When a merger or collision occurs between main-sequence stars, the remnant star is rejuvenated. We computed the effective age $A_{\rm MS,3}$ of the remnant star, i.e., the age of a single star whose remaining main‑sequence lifetime matches that of the merger product, according to Appendix C of \citealt{Vergara2025} (but see also, e.g., \citealt{Leonard1989,Lombardi2002,Hurley2002,Glebbeek2008,Mapelli2016,Schneider2016,Schneider2025,Rantala2026}):
\begin{equation}
    A_{\rm H , MS,3} = \frac{0.1}{1+q}\,\tau_{\rm MS,3}\,\left(\frac{A_{\rm MS,1}}{\tau_{\rm MS,1}} + q\,\frac{A_{\rm MS,2}}{\tau_{\rm MS,2}}\right),
\end{equation}
\begin{equation}
    A_{\rm MS,3} = A_{\rm MS,1} - q\,(A_{\rm MS,1} - A_{\rm H, MS,3}),
\end{equation}
where $q = m_2/m_1 < 1$, $A_{\rm MS,i}$ and $\tau_{\rm MS,i}$ are the effective ages and main-sequence lifetimes of the two colliding stars and the remnant star (with subscripts $i = 1,2$ referring to the primary and secondary stars, respectively, and $i = 3$ to the merger remnant). Note that this rejuvenation prescription contains explicitly the mass ratio between colliding stars. This parameter regulates the rejuvenation itself and indicates that when the star is very massive compared to the accreted stars its age is not substantially reduced anymore. We also note that the coefficient 0.1, which depends on the mass of the hydrogen core, is calibrated at $Z = 0.01$. At the lower metallicities considered here, this coefficient should be likely higher, possibly leading to a more efficient rejuvenation and extending the stellar lifetime by a few years. A more detailed assessment of the impact of rejuvenation will be explored in future works.

The mass loss during a stellar merger or collision is parametrized as:
\begin{equation}
\label{eq:floss}
f_{\rm loss} = D\,\frac{q}{(q+1)^2},
\end{equation}
where $D = 0.3$ \citep{Glebbeek2013}. The stellar lifetime and radius of the remnant are computed as previously described, where in this case $m_* = (m_1+m_2)\times(1-f_{\rm loss})$.

The primary star mass loss via stellar winds is treated following the prescription of \cite{Vink2018}, extended here also to stars with masses above $900\,\Ms$.  We did not consider mass loss from the secondary star or from other stars involved in dynamical interactions. Since the most massive stars are rapidly accreted through repeated mergers, we expect mass loss from stars other than the primary to play only a secondary role in the evolution.

The final fate of each star (i.e., whether it forms a BH and with what mass, or instead undergoes a PISN) is determined by the relation between the pre-supernova (pre-SN) stellar mass and the remnant mass derived from \sevn-\textsc{Parsec} tracks (see upper panel of Fig. \ref{fig:rem_mass_f_rad}). This choice helps to ensure consistency across different mass-loss histories. We adopted the rapid core‑collapse supernova formalism of \cite{Fryer2012}, while PISNe follow the prescriptions of \cite{Mapelli2020}. We also note that the onset of PISNe is primarily determined by the core mass rather than the pre-SN mass of the star. However, stellar collisions and the associated rejuvenation can alter the core growth, potentially leading to core masses that differ from those of single stars of the same ZAMS mass. Therefore, adopting pre-SN mass thresholds represents a conservative assumption.

We discuss the implications of our astrophysical assumptions in more detail in Appendices \ref{app:stellar_ev} and \ref{app:mass_loss}. Our prescriptions are broadly consistent with \cite{Rantala2026}, enabling comparison across models.

\renewcommand{\arraystretch}{1.6}
\begin{table}
\centering
\caption{Values of the fitting parameters.}
\label{tab1}
\begin{tabular}{l c c c}
\hline\hline
 & Parameter & $Z = 0.0002$ & $Z = 0.002$ \\
\hline
\multirow{2}{*}{$R_{\rm ZAMS}$ [R$_{\odot}$]} & $a_*$ & $0.757\pm 0.002$ & $0.763\pm 0.004$ \\
                         & $b_*$ & $0.553\pm 0.001$ & $0.587\pm 0.001$ \\
\hline
\multirow{3}{*}{$\tau_{\rm MS}$ [Myr]} & $a_{\rm MS}$ & $194\pm4 $ & $205\pm4$ \\
                         & $b_{\rm MS}$ & -1.138$\pm 0.005$ & -1.151$\pm 0.005$ \\
                         & $c_{\rm MS}$ & 1.752$\pm 0.006$ & 1.766$\pm 0.006$ \\
\hline
\multirow{3}{*}{$\tau_{\rm life}$ [Myr]} & $a_{\rm life}$ & $205\pm4$ & $213\pm  4$ \\
                         & $b_{\rm life}$ & -1.138$\pm 0.006$ & -1.146$\pm 0.006$ \\
                         & $c_{\rm life}$ & 1.954$\pm0.007$  & 1.951$\pm 0.006$ \\
\hline
\end{tabular}
\tablefoot{The first column indicates the stellar quantities (Eqs. \ref{eq:r_zams}, \ref{eq:t_evo}, and \ref{eq:tlife}, respectively), the second one lists the fitting parameters, and the third and fourth ones report their values for $Z = 0.0002$ and $Z = 0.002$, respectively.}
\end{table}
\renewcommand{\arraystretch}{1.0}

\subsection{Binary pairing and processes}
\label{sec:dynamics}
Observations indicate that massive stars mainly live in binaries or higher-order multiple systems \citep{Sana2012,Moe2017}. Their shorter mass-segregation timescales and larger cross sections compared to single stars imply that massive binaries play a central role in the dynamical evolution of star clusters. In addition, typically only one star per cluster can grow through runaway collision \citep{PortegiesZwart2002,FreitagI}. To model this scenario, we draw the primary mass from a flat distribution in the range $m_1 = 100 - 150 \,\Ms$, while the companion mass is determined by the binary mass ratio $q$. Following \cite{Moe2017}, binaries are grouped into three mass-ratio intervals, $(0-0.5)$, $(0.5-0.6)$, and $(0.6 - 1)$, containing approximately 20\%, 40\%, and 40\% of systems, respectively. One of these intervals is selected randomly based on the percentile, and the mass ratio is then drawn from a uniform distribution within the chosen range. The companion mass is finally set to $m_2 = \max(5\,\Ms\,,\, q\,m_1)$, where the minimum mass reflects observational constraints \citep{Sana2012,Kiminki2012,Kobulnicky2014}.

The binary eccentricity is drawn from a thermal distribution. The initial semi-major axis is extracted from a logarithmically flat distribution limited between the sum of the stars' radii and 50 AU. These assumptions permitted us to adopt a setup widely used in direct $N$-body simulations (see e.g., \citealp{Wang2016,Kamlah2022,DragonI}). 

Binary-single interactions are modeled via statistical treatment based on seminal works by \cite{Heggie1975,Hut1983,Heggie1993,Sigurdsson1993}. This formalism allows us to determine the outcome of the interactions (i.e., hardening or softening, dynamical exchange, binary ionization, or recoil and subsequent ejection from the cluster) based on energy considerations during the encounter. All possible cases are discussed in detail in Appendix \ref{app:bs_int}. 

Here, we summarize the formalism we employed to model these interactions. If one considers a binary with mass $m_{\rm bin} = m_1+m_2$, semi-major axis $a$, and eccentricity $e$, traveling through a population of objects characterized by mean mass $m_{\rm mean}$, number density $n$, and velocity dispersion $\sigma$, the interaction rate between the binary and a single object in such an environment can be written as:
\begin{equation}
    \frac{{\rm d}n_{ \rm bs}}{{\rm d}t} = n\, \sigma \,\Sigma_{\rm bs},
\end{equation}
where $\Sigma_{\rm bs}$ is the binary cross-section:
\begin{equation}
    \Sigma_{\rm bs} = \pi\, a^2\, (1-e)^2 \left[1+\frac{2\,G\,(m_{\rm bin}+m_{\rm mean})}{\sigma^2\, a\,(1-e)}\right],
\end{equation}
the term in the square brackets, accounting for the gravitational focusing, is largely greater than 1 in most applications. The associated timescale of this process is given by $t_{\rm bs} = \left(n\,\sigma\,\Sigma_{\rm bs}\right)^{-1}$ \citep{Miller2009,Antonini2016}.

On the other hand, the collision cross-section is computed by assuming that the pericenter distance equals the stellar radius:
\begin{equation}
\Sigma_{\rm coll} = \pi \,(R_{\rm i} + R_{\rm p})^2 \left[1+\frac{2\,G\,(m_{\rm i}+m_{\rm p})}{\sigma^2 \, (R_{\rm i} + R_{\rm p})}\right],
\end{equation}
where $m_{\rm i}$ and $R_{\rm i}$ denote the mass and radius of the binary component, while $m_{\rm p}$ and $R_{\rm p}$ are the mass and radius of the perturber. We deepen this assumption in Appendix \ref{app:contact}. The velocity dispersion and number density adopted here correspond to those of the cluster core. To determine which binary component undergoes the collision, we compared their gravitational focusing terms; the component with the larger gravitational focusing is assumed to collide. The corresponding collision timescale, $t_{\rm coll}$, is then computed following the same procedure as for the binary–single interaction case. When a collision occurs, we neglect any recoil kicks. While this may be a strong assumption, hydrodynamical simulations \citep{Davies1991,Lombardi2002,Ballone2023} indicate that recoil velocities are typically $\lesssim \sigma$. Furthermore, as the star grows in mass, the recoil velocity is expected to become progressively smaller. However, a binary may be ionized following a collision if the kinetic energy of the perturber exceeds the binding energy of the binary. The velocity of the perturber is sampled from a Maxwellian distribution with dispersion $\sigma$. 

After an interaction, if the new binary pericenter is smaller than the sum of the two stars' radii, the stars are merged and the merger remnant is possibly paired with a new companion. If the stellar merger occurs during a resonant interaction, the merger remnant is paired with the third body and the newly formed binary is assumed to have a semi-major axis $a_{\rm new} = (m_3 / m_2)a_{\rm old}$, with $m_2 < m_1$ \citep{Sigurdsson1993}. Otherwise, after a merger or ionization, we assumed that the remnant star will be captured in another binary over a time $t_{\rm bin} \equiv t_{\rm 3bb}$ \citep{Lee1995}:
\begin{equation}
\label{eq:t3bb}
    \begin{split}
          t_{\rm 3bb} \sim 48.6\text{ Myr}\, \left(\frac{v_*}{10\,\rm km\, s^{-1} }\right)^9\,\left(\frac{m_{\rm mean}}{20\, \text{M}_{\odot}}\right)^{-5}\,\times\\
          \times\,\left(\frac{n}{10^5 \,\text{pc}^{-3}}\right)^{-3}\,\left(\frac{R_{\rm cl}}{0.1 \,\text{pc}}\right)^{-3},  
    \end{split}
\end{equation}
where $v_* \sim \sigma$, $R_{\rm cl}$ is the radius of the cluster, and $n$ its number density.
The binary formed this way is assigned a semi-major axis from a logarithmically flat distribution between $a_{\rm new} = (10^{-2}-10)\,a_{\rm h}$, $a_{\rm h} = G\,m_{\rm mean}/4\,\sigma^2$ \citep{Sesana2006}, which are typical values observed in self-consistent $N$-body models, and an eccentricity extracted from a thermal distribution. 

In \starfall, the scheme above is implemented following a multi-stepped strategy (see Fig. \ref{fig:flowchart}). We assume that the cross section and the corresponding interaction timescale (either $t_{\rm bs}$ for a binary–single interaction if the binary remains bound or $t_{\rm coll}$ for a possible collision) quantify the probability of such an interaction occurring in a time interval $dt$. In particular, at each new time-step $dt$, the timescales $t_{\rm bs}$ and $t_{\rm coll}$ are evaluated. Then,
\begin{itemize}
    \item if $t_{\rm bs}< \min(t_{\rm coll}\,,\,dt)$, the system undergoes a binary–single interaction, with the outcome (e.g., hardening, dynamical exchange, or ionization) determined following \cite{Sigurdsson1993}, as described in Appendix \ref{app:bs_int};
    \item if $t_{\rm coll} < \min(t_{\rm bs}\,,\, dt)$, a collision event is assumed to occur;
    \item if the binary is ionized and $t_{\rm coll} < dt$, the single star can continue to experience collisions until a new binary is formed;
    \item  multiple interactions may occur within a single $dt$; after each interaction, $t_{\rm coll}$ and $t_{\rm bs}$ are re-evaluated, and the time is advanced by $\min(t_{\rm coll},t_{\rm bs})$ until the cumulative elapsed time exceeds $dt$; 
    \item if both $t_{\rm coll}$ and $t_{\rm bs}$ exceed $dt$, an interaction may occur with a probability $P_{\rm int} = dt/\min(t_{\rm coll},t_{\rm bs})$;
    \item if no interaction takes place during a timestep, the simulation advances to the next one.
\end{itemize}

If an interaction occurs, we extract the mass of the third object from a set of $N_{\rm core} = M_{\rm core}/m_{\rm mean}$ masses distributed as a Kroupa initial mass function (IMF; \citealt{Kroupa2001}), between $0.08-150 \,\Ms$, with $m_{\rm mean} = 0.6\,\Ms$. Here, $N_{\rm core}$ sets the reservoir of stars available for interactions, while the actual number of stars present at each timestep is updated consistently according to the evolving core mass. We assumed that in case of any interaction, the binary will interact more likely with the most massive stars. Hence, we initially drew the ZAMS mass of the perturber in the range $m_3 = 20-150\,\Ms$. In case of a merger or collision, we removed the involved object (which can be either one of the binary components or the perturber) from the pool of stars in the core. Once all massive stars are consumed or have already collapsed, perturbers are drawn from the remaining lower-end of the IMF. Our analysis considers only stars that are initially present in the core. The impact of mass segregation will be investigated in future works. It is nevertheless expected to have a secondary effect, as it occurs on timescales $\gtrsim 1$ Myr for a star with $m_* = 50 \,\Ms$ \citep{PortegiesZwart2002} in the most compact clusters, i.e., after the collisional runaway has already been triggered, or even on longer timescales in less compact environments.

All simulations are halted when the primary star collapses.


\begin{figure*}
    \centering
    \includegraphics[width=\textwidth]{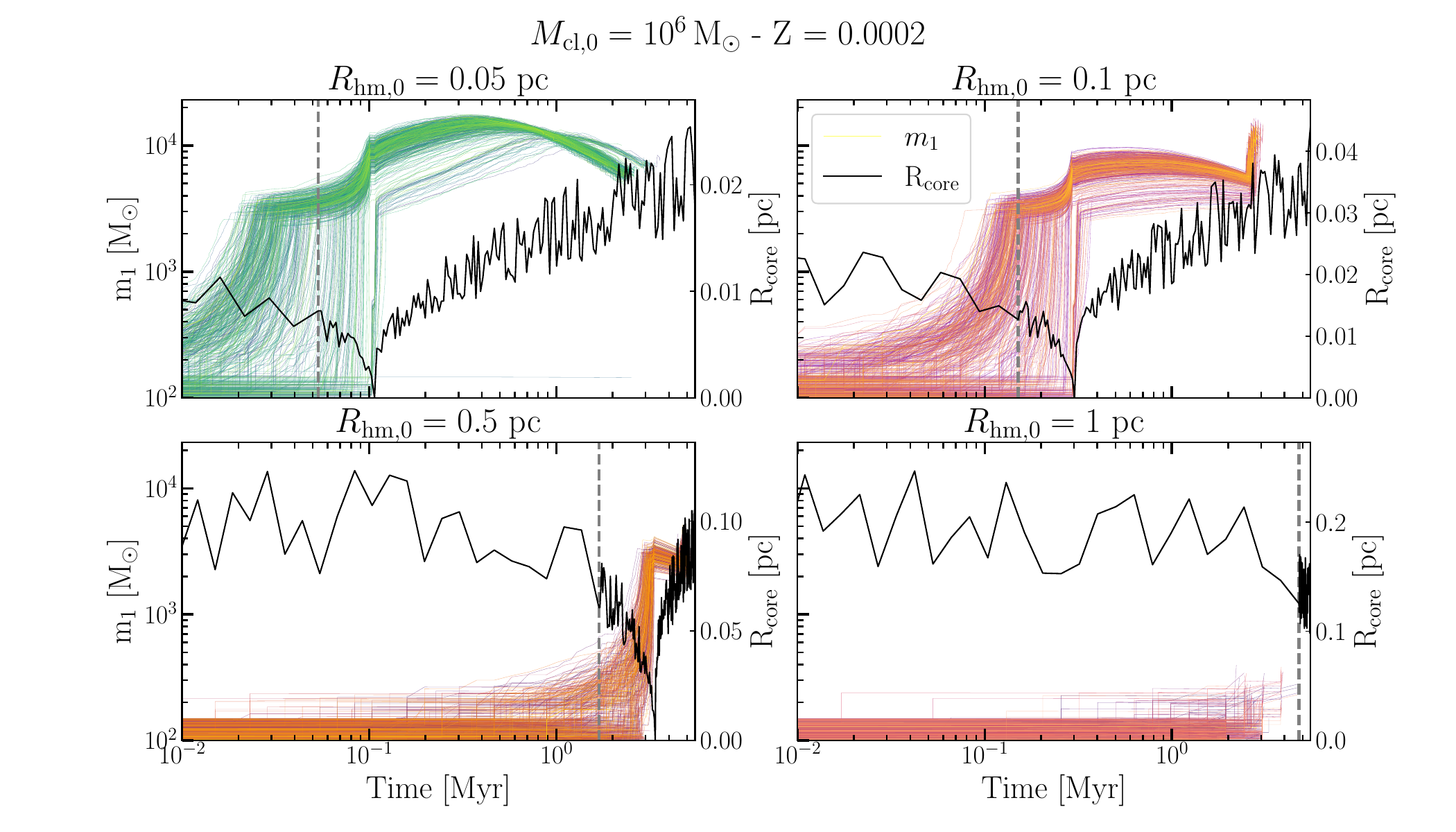}
    \caption{Mass growth of VMSs (colored lines) in clusters of $M_{\rm cl,0} = 10^6 \,\Ms$ and $R_{\rm hm,0} = 0.05, 0.1,0.5,1$ pc. Black lines represent the evolution of the cluster core radius. The vertical gray dashed lines mark the beginning of the core-collapse phase.}
    \label{fig:vms_mass_growth}
\end{figure*}

\section{Results}
\label{sec:results}

\subsection{VMS mass growth}

The onset and extent of the collision chain vary with the cluster’s properties, the frequency of dynamical interactions, and the evolutionary state of the primary star. Figure \ref{fig:vms_mass_growth} shows the growth of the primary star mass and the evolution of the cluster core radius in clusters with total mass $M_{\rm cl,0} = 10^6  \,\Ms$ for selected half-mass radii. In the two densest clusters ($R_{\rm hm,0} = 0.05,0.1$ pc), the star can begin growing mass even before core collapse, followed by a steeper mass increase during the core-collapse phase. After this phase, the mass gained through dynamical interactions becomes smaller than the mass lost via stellar winds, leading to a gradual decrease in stellar mass until the end of its life at $\sim 3$ Myr. In both clusters, core depletion leads to the formation of VMSs with masses of $\sim 10^4\,\Ms$, but on different timescales. In the first cluster, core collapse occurs before the onset of post-main-sequence stellar expansion. By the time the star expands, the cluster core is already depleted. As a result, the increased stellar radius does not lead to additional collisions or a further increase in the maximum mass. In the second one, instead, stellar expansion drives further mass growth phase before the direct collapse. However, in this case, the resulting puffed-up structure may lead to enhanced collision-induced mass loss compared to the estimates of \cite{Glebbeek2013} adopted in this work. In contrast, considering $R_{\rm hm,0} = 0.5$ pc, we observe only a few mergers/collisions before the onset of core collapse. In this case, the core-collapse time is roughly equal to $\tau_{\rm MS}$, i.e., the time after which the star expands (see Sect. \ref{sec:stellar_evolution} and Eq. \ref{eq:t_evo}). The increased core density and the large star's radius maximize the interaction rate, driving a fast increase of the VMS mass. The delayed mergers rejuvenate the star efficiently over a longer timescale, prolonging its life up to $\sim$5.2 Myr. In less dense clusters, the lower interaction rate prevents, in most cases, a runaway phase, with only few stars exceeding 300 $\Ms$.

\begin{figure}
    \includegraphics[width=\columnwidth]{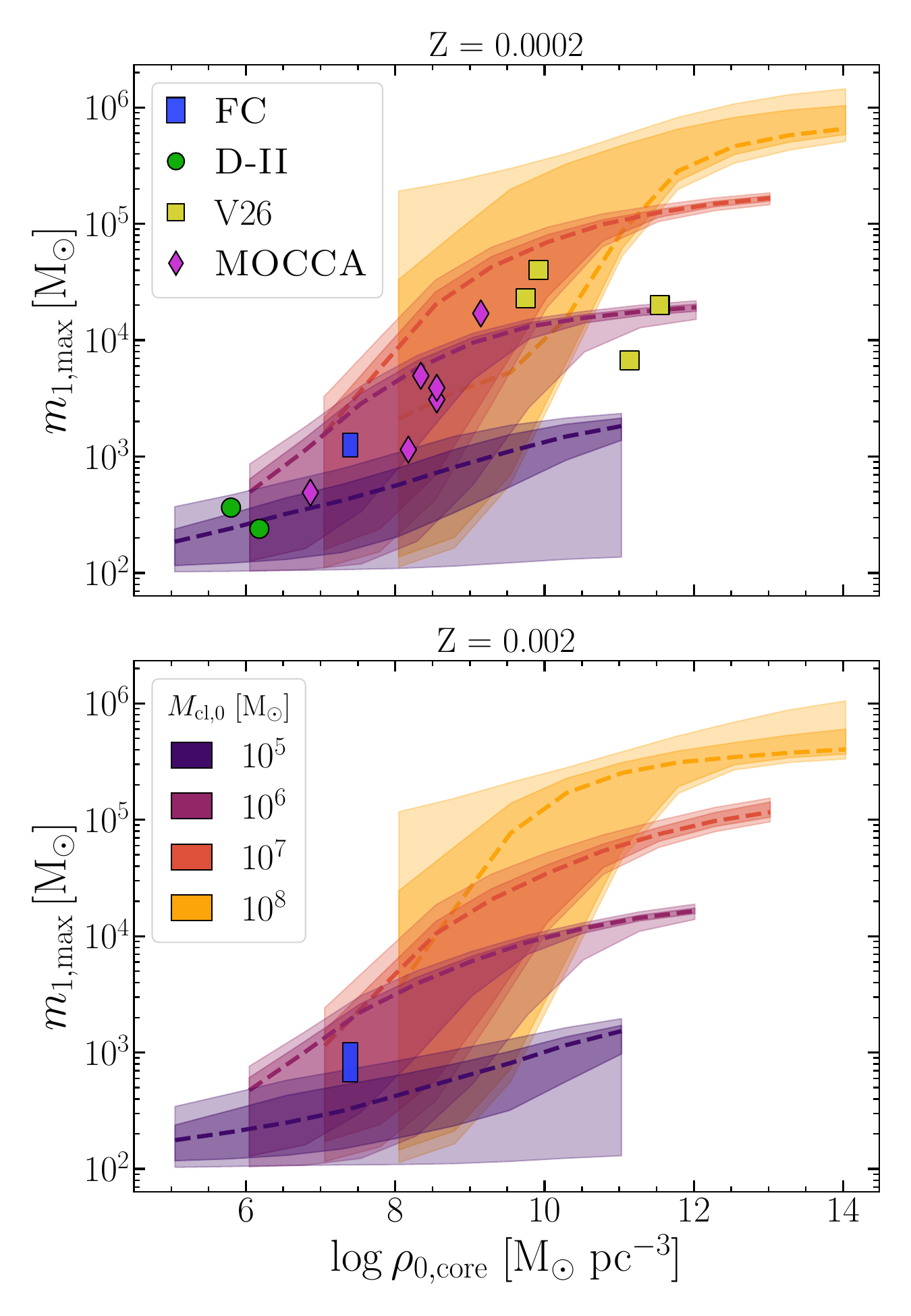}
    \caption{Median (dashed lines) mass at the end of the collision chain of VMSs as a function of the initial central density for $Z = 0.0002$ (upper panel) and $Z = 0.002$ (lower panel). The colors encode the initial mass of the cluster, from $M_{\rm cl,0} = 10^5 \,\Ms$ (violet) to $M_{\rm cl,0} = 10^8 \,\Ms$ (yellow). The dark (light) shaded region encloses 50\% (90\%) of the entire population. The blue area indicates the range of stellar masses reported by \citealt{Rantala2026} (\textsc{Frost-clusters}; \textsc{FC}), the green dots mark those reported by \citealt{DragonII} (\textsc{Dragon-II}; \textsc{D-II}), the yellow squares those by \citealt{Vergara2026} (\textsc{Nbody6++GPU} and \textsc{MOCCA}; \textsc{V26}), while the purple diamonds those by Askar et al. in prep, Giersz et al. in prep (\textsc{MOCCA}).}
    \label{fig:stat_mass}
\end{figure}

Figure \ref{fig:stat_mass} shows the statistical properties of the maximum mass reached by VMSs. A comparison between the two panels highlights the key roles of stellar dynamics. We find a clear correlation between the maximum VMS mass and both the initial core density and the total mass of the cluster. Metallicity primarily affects the width of the mass distribution and limits the maximum mass by a few percent, especially for clusters with $M_{\rm cl,0} \geq 10^6\,\Ms$.

At low cluster mass ($M_{\rm cl,0} = 10^5 \,\Ms$), the VMS mass remains relatively small, in the range $\sim 100-1000\,\Ms$. In more massive clusters (e.g., $M_{\rm cl,0} = 10^8 \,\Ms$), the maximum mass can exceed $10^4\,\Ms$ and grow up to $\sim 10^6\,\Ms$ for clusters with $M_{\rm cl,0} = 10^8\,\Ms$. Such extremely massive stars may undergo general relativistic instabilities that trigger the collapse of their inner core, leading to the formation of a very massive BH surrounded by a stellar envelope feeding the accretion disk \citep{Hosokawa2013}. Our current implementation does not include a dedicated prescription to model this phase. Therefore, for simplicity, these objects are treated as ordinary single stars.  While the total cluster mass sets the overall reservoir of stars available for collisions and dynamical interactions, the core density determines the interaction rate, and therefore how much of that reservoir can be accreted. The central density, indeed, regulates how efficiently and how rapidly the VMS can grow. As the initial core density increases, the maximum VMS mass rises accordingly, reflecting the enhanced frequency of close encounters and mergers. However, for $\rho_{\rm core,0} \gtrsim 10^{9}\,\Ms\,{\rm pc}^{-3}$, the growth saturates and both the 50\% and 90\% intervals converge toward the median. In this regime, the primary star effectively accretes all the mass available in the cluster core.

To assess the reliability of our approach, we compared our predictions with results from direct $N$-body \citep{ArcaSedda2023, Vergara2026, Rantala2026} and Monte Carlo (\citealt{Askar2025}; Askar et al. in prep; Giersz et al. in prep) simulations. This comparison allows us to evaluate how well our semi‑analytical approach captures the key physical processes driving runaway growth. Despite the simplifying assumptions adopted in our model, our predictions are in overall good agreement with the $N$-body results of \citealt{DragonII} ($M_{\rm cl,0} = 0.7\times10^5 \,\Ms$, $R_{\rm hm,0} = 0.7\,,\,1.75$ pc, $Z = 0.0005$) and \citealt{Rantala2026} ($M_{\rm cl,0} = 1.5\times10^5 \,\Ms$, $R_{\rm hm,0} = 0.37$ pc, $Z=0.0002,0.002$), as well as Monte Carlo simulations by Askar et al. in prep., Giersz et al. in prep. ($M_{\rm cl,0} = (1.3-6.7)\times10^6 \,\Ms$, $R_{\rm hm,0} = 0.25 - 3.3$ pc, $Z = 0.0005,0.0006$), for clusters with masses and half‑mass radii comparable to those explored in this work. On the other hand, the masses reported by \citet{Vergara2026}, both with $N$-body and Monte Carlo simulations, are up to an order of magnitude higher than our estimates for clusters with $M_{\rm cl,0} = 10^6 \,\Ms$. Their simulations focused on clusters with $M_{\rm cl,0} = (0.58-4.38)\,\times\,10^5 \,\Ms$, $R_{\rm hm,0}=0.01-0.05$ pc and $Z = 0.0001$. As also noted by \cite{Rantala2026}, significant differences may arise from the adopted prescriptions for stellar‑wind mass loss and stellar radii. In this work, we employed the wind prescription from \citet{Vink2018}, which leads to more substantial mass loss in VMSs than the prescription of \citet{Vink2001}, implemented by \cite{Vergara2026} with an upper limit of $1.5\times10^{-4}\,\Ms\,\mathrm{yr}^{-1}$ for stars exceeding $6\times10^{5}\,\mathrm{L}_{\odot}$ \citep{Humphreys1994,Belczynski2010}. Nevertheless, even with these stronger winds our model predicts the formation of stars with masses exceeding $10^3 \,\Ms$, largely consistent with IMBH progenitors (see also Appendix~\ref{app:mass_loss} for a detailed assessment of the impact of stellar winds). In addition, radii from \sevn-\textsc{Parsec} tracks are smaller with respect to the prescription implemented in BSE (see, e.g., \citealt{Hurley2002,Hurley2005}). Other differences may stem from the underlying cluster models and, consequently, the mass of their core. This number may be enhanced by rapid, or even primordial, mass segregation. We leave a thorough exploration of the impact of such a mechanism to a forthcoming work. 

Table \ref{tab2} dissects the maximum VMS mass in our models into different ranges of cluster mass ($M_{\rm cl,0}$) and radius ($R_{\rm hm,0}$). The table further highlights the tight relation between cluster mass, density, and terminal VMS mass. Cluster models with sizes and masses comparable to the most massive GCs in the Milky Way can nurture the growth of a VMS as massive as 900 $\,\Ms$.

\begin{figure}[!ht] 

    \includegraphics[width=\columnwidth]{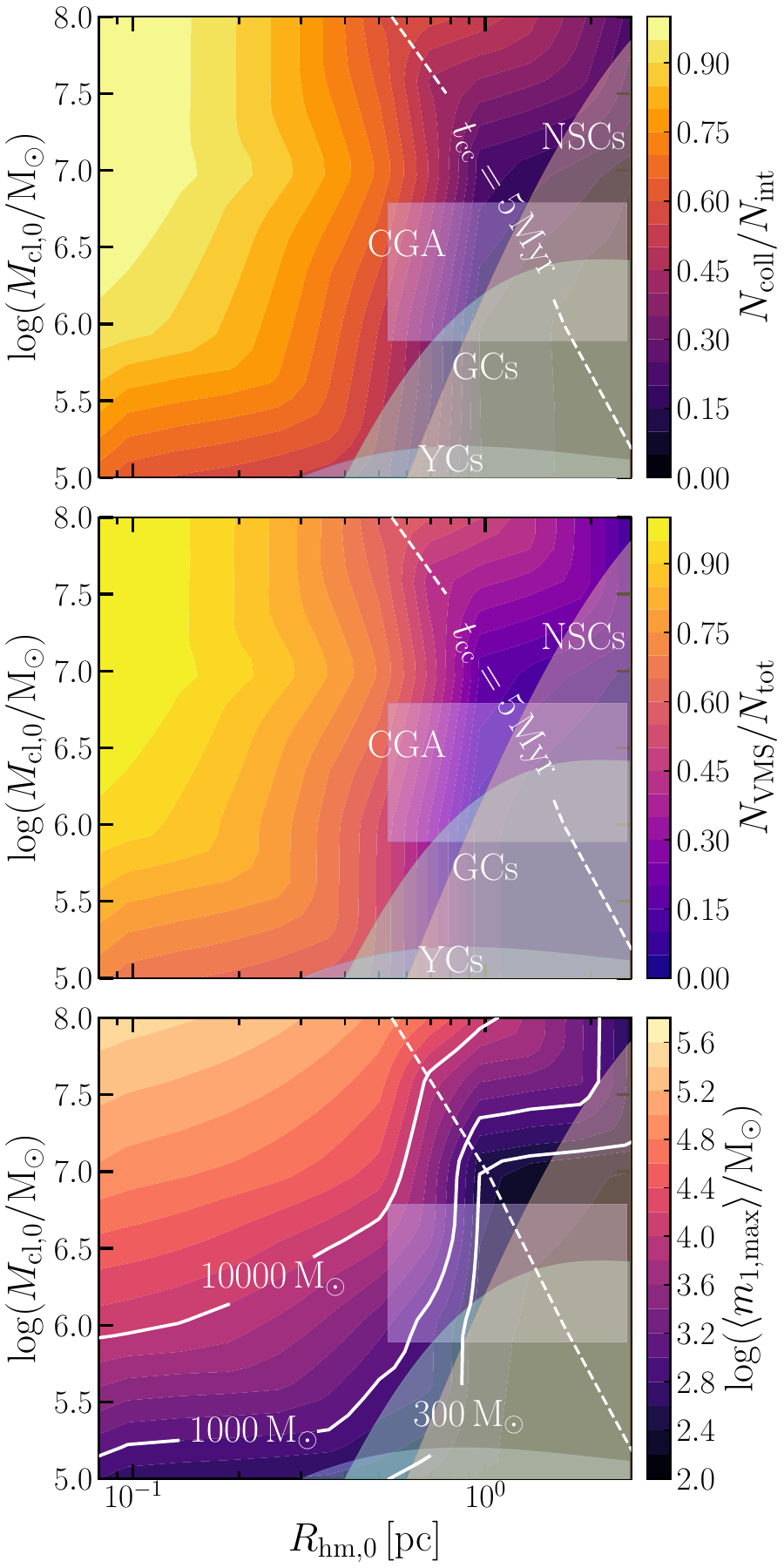}
    \caption{Contour plots in the $M_{\rm cl}$–$R_{\rm hm}$ plane at $Z = 0.002$ showing, from top to bottom: the number of collisions over the total number of interactions, $N_{\rm coll}/N_{\rm int}$; the number of VMSs with mass $m_* \geq 300\,\Ms$ over the total number of simulated binaries, $N_{\rm VMS}/N_{\rm tot}$; the mean maximum stellar mass. In all panels, the shaded regions indicate the typical ranges of $R_{\rm hm}$ and $M_{\rm cl}$ for clusters in the Cosmic Gems Arc (CGA), young clusters (YCs) from \cite{Messa2026}, and globular clusters (GCs), and nuclear star clusters (NSCs) from \cite{Paiella2026}. The dashed line marks clusters with a core-collapse timescale $t_{\rm cc} = 5$ Myr. All cluster properties are considered at the present known age.}
    \label{fig:contour_coll_vms_growth}
\end{figure}

As detailed in the previous section, in \starfall\ the VMS can grow either via merger of binary components, hierarchical collision, or their combination. In order to assess the mutual role of such processes in the VMS buildup, Figure \ref{fig:contour_coll_vms_growth} shows how cluster mass and initial radius affect the fraction of stellar collisions, fraction of stars formed with $m_* > 300 \,\Ms$, and mean VMS mass. We overlay typical observed properties of Galactic GCs \citep{Harris2015}, nuclear star clusters (NSC; \citealt{Georgiev2016}), young massive clusters (YC; \citealt{PortegiesZwart2010}), as reported by \cite{Paiella2026}, and the young clusters observed in the CGA at a redshift $z\sim10$ \citep{Adamo2024,Messa2026}. To keep the figure readable, the young and massive clusters present in the galaxy Henize 2-10 (with $M_{\rm cl} \sim (0.2-2.3)\times10^6\,\Ms $ and $R_{\rm hm} \sim 1.2-4$ pc; \citealt{Reines2011,Nguyen2014}) were not included. However, they are consistent with the most massive GCs and lower-mass clusters in the CGA. Therefore, the discussion that follows also applies to the clusters in Henize 2-10. The plot highlights that stellar collisions are more efficient than stellar mergers at producing more massive VMS in the most massive and compact clusters. Indeed, a higher stellar collision efficiency is strictly correlated with a higher probability of VMS formation. Typically, the main bottleneck for triggering a collision runaway is the merger of the initial binary. In the densest systems ($R_{\rm hm,0} < 1$ pc and $M_{\rm cl,0} > 10^6 \,\Ms$), once this first merger occurs, the remnant undergoes a prolonged sequence of collisions without forming a new binary. As the central density decreases, stellar mergers become the main trigger of VMS formation, but dynamical ejection and smaller interaction rates reduce the attainable mass, leading to lighter VMS. In addition, a VMS forms in $\sim15\%-50\%$ of the models even for $t_{\rm cc} \gtrsim 5$ Myr (a typical time threshold for the onset of runaway collisions; see, e.g., \citealt{PortegiesZwart2002}), provided that the cluster is sufficiently massive ($M_{\rm cl} \gtrsim 10^{7}\,\Ms$). In systems with $R_{\rm hm,0} \gtrsim 1$ pc, dynamics do not play a significant role, and stars with $m_* > 300\, \Ms$ are formed only in the most massive clusters (i.e., $M_{\rm cl,0}> 10^7\,\Ms$). Comparing our models to observed clusters suggests that massive and dense GCs were able to support the growth of VMSs of $\gtrsim$ 300 $\,\Ms$, and probably beyond 1000 $\Ms$. Indeed, since these old systems likely underwent significant mass loss and expansion over their lifetime, their mass - radius distribution was originally likely skewed upward and leftward in the $M_{\rm cl,0}-R_{\rm hm,0}$ plane \citep{Baumgardt2018,Bianchini2026}, i.e. in a region of the phase-space where even heavier VMSs could form. Interestingly, some of the CGA clusters seem to fall exactly in such region, where VMS mass can reach $10^3-10^4 \,\Ms$. Our results are consistent with recent findings from \cite{Williams2026}, who developed an analytic cosmological framework to explore the dynamical evolution of dense star clusters with and without a central massive BH. In their analysis, \cite{Williams2026} find that stellar collisions in sufficiently dense environments can lead to the rapid growth of VMS in high-redshift and compact clusters, reaching masses consistent with the one we find in \starfall, despite the different adopted approach. We further discuss this aspect in the next section.

\begin{figure*}[!ht] 
    \centering
    \includegraphics[width=0.9\textwidth]{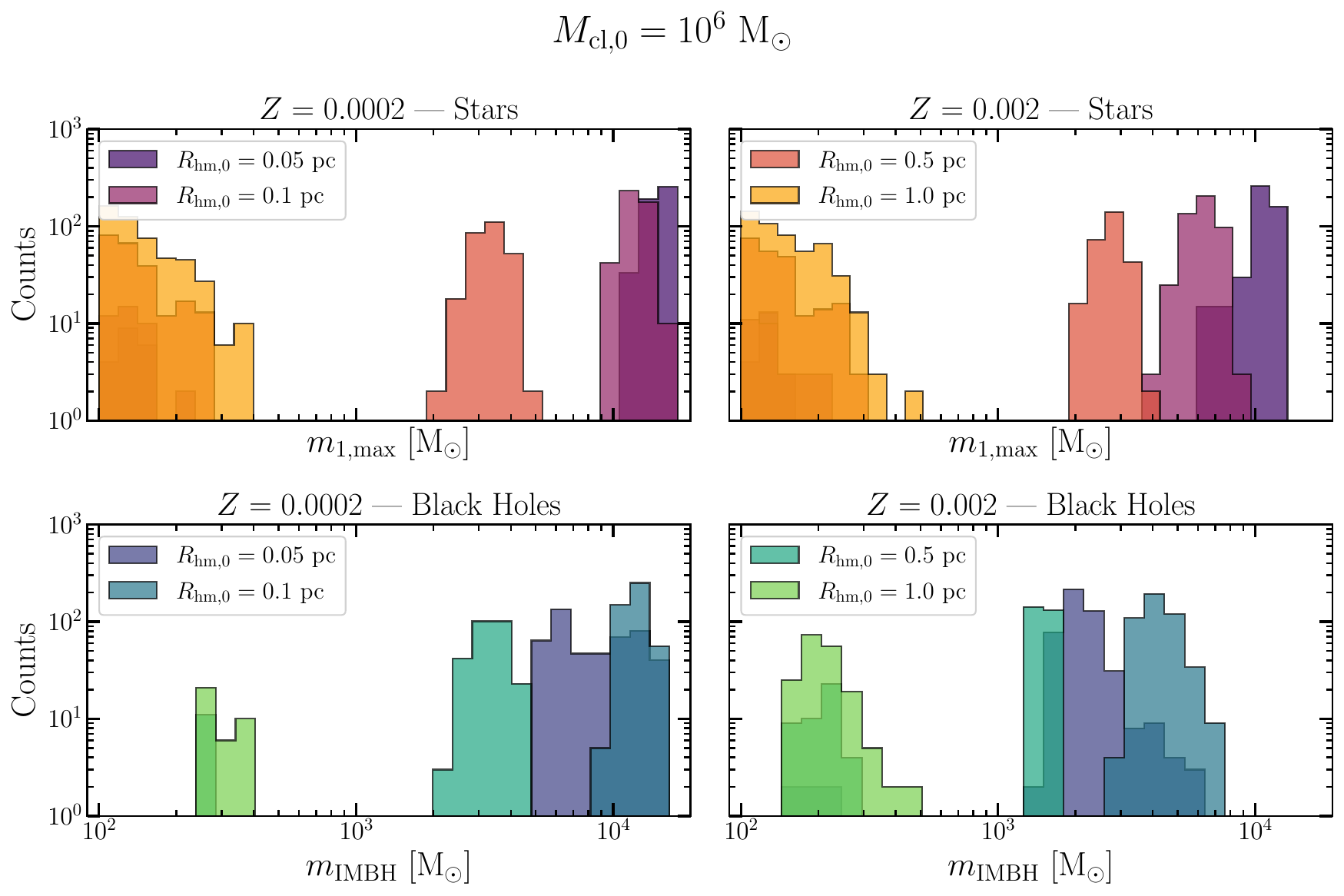}
    \caption{Stellar mass distribution (upper panels) and IMBH mass distribution (lower panels) for clusters with mass $M_{\rm cl,0} = 10^6\,\Ms$ and half-mass radius $R_{\rm hm,0}$ in the range $0.05$–$1$ pc. The left (right) panels correspond to metallicity $Z = 0.0002$ ($Z = 0.002$).}
    \label{fig:histo_vms_imbh}
\end{figure*}

\begin{figure}[!ht]
    \includegraphics[width=\columnwidth]{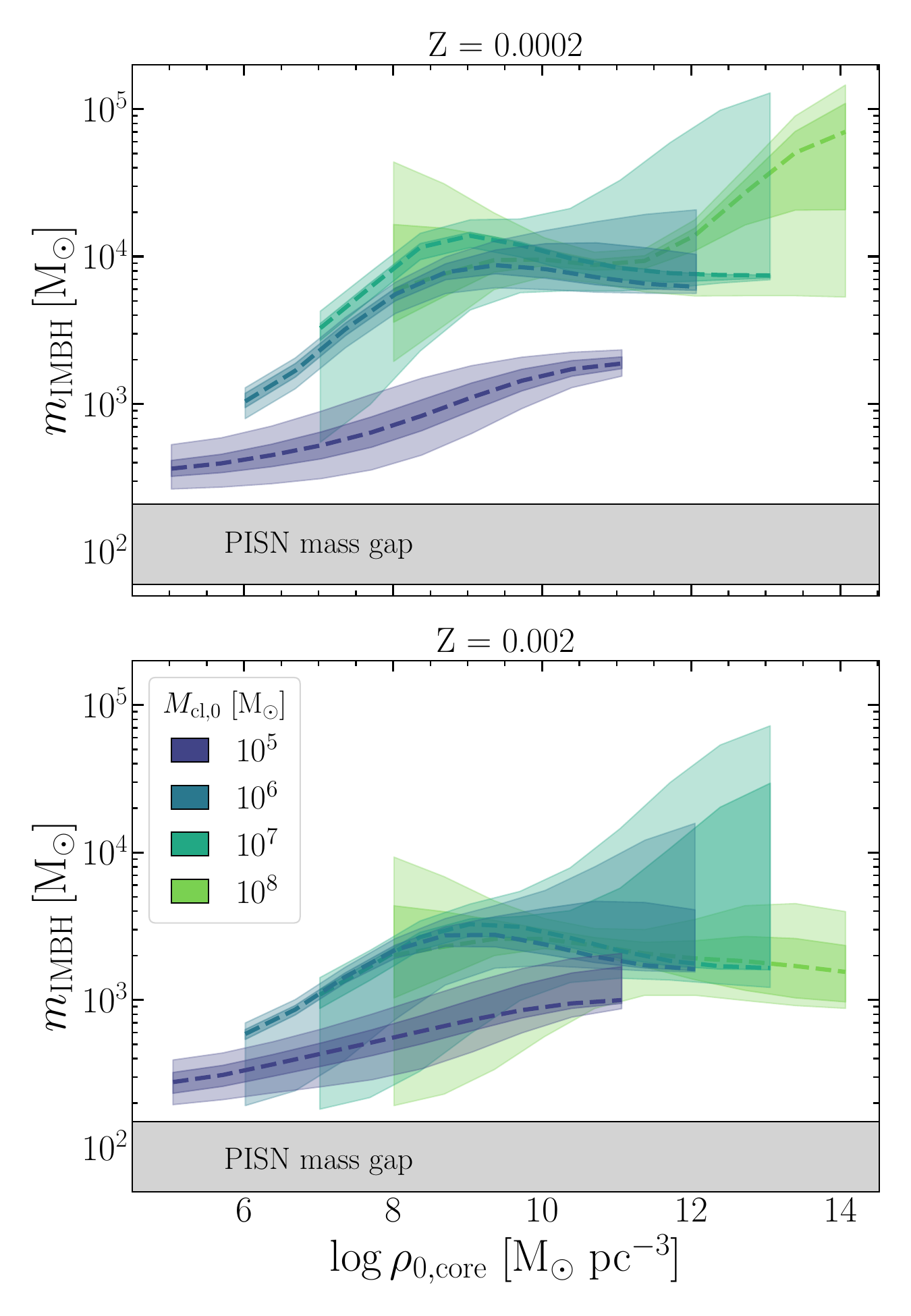}
    \caption{Same as Fig. \ref{fig:stat_mass}, but for IMBH masses. The colors encode the initial mass of the cluster, from $M_{\rm cl,0} = 10^5 \,\Ms$ (violet) to $M_{\rm cl,0} = 10^8 \,\Ms$ (light green). The dark (light) shaded region encloses 50\% (90\%) of the entire population. The gray band highlights the PISN mass gap for the two metallicities.}
    \label{fig:stat_imbh}
\end{figure}

\begin{table} 
\centering 
\setlength{\tabcolsep}{6pt} \renewcommand{\arraystretch}{1.9} 
\caption{Median maximum stellar mass in different star clusters.} 
\label{tab2} 
\begin{tabular}{c c | c c c} 
\hline\hline 
& & \multicolumn{3}{c}{$R_{\mathrm{hm},0}$ [pc]}\\ 
& & $0.1 - 0.5$ & $0.5 - 1$ & $1 - 3$\\ 
\hline 
\multirow{3}{*}{\rotatebox{90}{$\log(M_{\mathrm{cl},0}/\mathrm{M}_{\odot})$}} & $5 - 6$ &$900^{+11\,000}_{-440}$          & $330^{+260}_{-190}$              & $127^{+19}_{-12}$ \\
& $6 - 7$ & $16\,000^{+56\,000}_{-3\,500}$   & $3\,200^{+17\,000}_{-3\,000}$    & $145^{+56}_{-23}$\\ 
& $7 - 8$ &  $108\,000^{+240\,000}_{-29\,000}$ & $29\,000^{+21\,000}_{-28\,000}$ & $189^{+120}_{-59}$ \\ 
\hline 
\end{tabular} 
\tablefoot{Median maximum stellar mass ($m_{\rm 1, max}$) as a function of initial cluster half-mass radius (columns; $R_{\rm hm,0}$) and cluster mass (rows; $\log(M_{\rm cl,0}/\,\Ms)$) at $Z = 0.0002$. The $\pm$ values give the 1$\sigma$ spread. The masses are given in $\,\Ms$ with the respective uncertainties.} 
\end{table}

\subsection{IMBH seeds}
\label{sec:imbh_seeds}

Depending on both the mass and the metallicity, a VMS may either undergo a PISN or directly collapse into an IMBH. Table \ref{tab3} reports the fraction of models that successfully form an IMBH seed. For clusters with initial mass $M_{\rm cl,0} = 10^6\,\Ms$ at metallicity $Z = 0.0002$ ($Z = 0.002$), the formation of IMBH seeds occurs in only $\sim 8\%$ ($\sim 37\%$) of the models for $R_{\rm hm} = 1$ pc, while this fraction increases to $\sim 96\% \,(96\%)$ for clusters with $R_{\rm hm} = 0.05$ pc. These cases correspond to the IMBH mass distributions shown in Figure \ref{fig:histo_vms_imbh}. VMSs that do not collapse into an IMBH, given our initial conditions, can either be disrupted in a PISN or undergo a pulsational pair-instability supernova (PPISN), forming a stellar mass BH. The maximum stellar mass and IMBH seed mass distribution for the other clusters we considered is presented in Appendix \ref{app:imbh_seed}. We discuss instead the impact of our astrophysical assumptions and the stellar mass BH distribution in Appendix \ref{app:stellar_ev}, while the impact of the mass-loss prescriptions we adopted is deepened in Appendix \ref{app:mass_loss}. 

The distribution of the maximum stellar mass reached at the end of the collision sequence is not significantly affected by metallicity. In both cases, the distribution spreads between $\sim100 \,\Ms- 4000\,\Ms$ ($\sim100 \,\Ms-400\,\Ms$) for $R_{\rm hm,0} = 0.5$ pc ($R_{\rm hm,0} = 1$ pc), while it is peaked at $\sim 10^4\,\Ms$ for $R_{\rm hm,0} = 0.05\,,\,0.1$ pc.

In contrast, both metallicity and the formation time of the VMS play an important role in determining the mass of the final remnant. In the densest clusters, the VMS forms earlier and rapidly reaches its maximum mass, above which stellar wind mass loss dominates over stellar collisions (see also Fig. \ref{fig:vms_mass_growth}). In less dense clusters, the VMS typically forms on longer timescales and therefore has less time to lose mass before collapsing, producing either more massive BHs or BHs with masses comparable to those produced in denser environments. In particular at $Z = 0.002$, while the maximum stellar mass reached during the collision sequence strongly depends on the dynamical properties of the cluster, the IMBH mass is primarily determined by the subsequent stellar evolution. Therefore, although the densest clusters produce the most massive stars, they do not necessarily produce the most massive IMBH seeds. Indeed, as shown in Fig. \ref{fig:histo_vms_imbh}, the maximum stellar mass reaches at most $m_* \sim 2\times10^4\,\Ms$ in the densest cluster ($R_{\rm hm,0} = 0.05$ pc), while the largest IMBH mass, $m_{\rm IMBH} \sim 1.5\times10^4\,\Ms$, is obtained at lower densities ($R_{\rm hm,0} = 0.1$ pc). This trend is particularly evident at $Z = 0.002$, due to the higher efficiency of stellar wind mass loss. In this case, the IMBH mass distribution for $R_{\rm hm,0} = 0.05$ pc peaks at $2\times10^3\,\Ms$, comparable to that for $R_{\rm hm,0} = 0.5$ pc, whereas the distribution for $R_{\rm hm,0} = 0.1$ pc peaks at $4\times10^3\,\Ms$.  At $R_{\rm hm,0}=1$ pc, instead, the conditions are not always favorable for triggering a runaway collision chain. In this case, the maximum stellar mass distribution exhibits a peak at $100-200\,\Ms$, corresponding approximately to the initial stellar masses and therefore indicating negligible mass growth. These stars do not contribute to IMBH formation, as they are expected to undergo PISNe. A tail extending beyond $200\,\Ms$ remains, increasing the probability of IMBH formation to up to $\sim60\%$ in more massive clusters (see Tab. \ref{tab3}).

While the probability of IMBH formation does not show any significant dependence on metallicity in the densest clusters, at $R_{\rm hm,0} = 1$ pc IMBH formation appears to be more efficient at $Z = 0.002$ (see also Tab. \ref{tab3}). At that metallicity many IMBHs are formed around a few hundred solar masses while the same IMBH subpopulation disappears at Z = 0.0002 (Fig. \ref{fig:histo_vms_imbh}). This can be explained by the effects of PISNe, which prevent stars with pre-SN masses $m_{\rm pre-SN} \lesssim 240\,\Ms$ at $Z = 0.0002$ from collapsing into an IMBH (see also Fig. \ref{fig:rem_mass_f_rad}).

\begin{table}
\centering
\setlength{\tabcolsep}{10pt}
\renewcommand{\arraystretch}{1.9}
\caption{Models forming an IMBH.}
\label{tab3}
\begin{tabular}{c c | cccc}
\hline\hline
& & \multicolumn{4}{c}{$R_{\mathrm{hm},0}$ [pc]} \\
& & 0.05 & 0.1 & 0.5 & 1 \\
\hline

\multicolumn{6}{c}{Z = 0.0002} \\
\hline
\multirow{4}{*}{\rotatebox{90}{$\log(M_{\mathrm{cl},0}/\mathrm{M}_{\odot})$}}
& $5$ & 78\% & 67\% & 57\% & 6\% \\
& $6$ & 96\% & 92\% & 57\% & 8\% \\
& $7$ & 99\% & 99\% & 71\% & 30\% \\
& $8$ & 96\% & 99\% & 68\% & 66\% \\
\hline

\multicolumn{6}{c}{Z = 0.002} \\
\hline
\multirow{4}{*}{\rotatebox{90}{$\log(M_{\mathrm{cl},0}/\mathrm{M}_{\odot})$}}
& $5$ & 78\% & 72\% & 55\% & 24\% \\
& $6$ & 96\% & 95\% & 65\% & 37\% \\
& $7$ & 99\% & 99\% & 85\% & 59\% \\
& $8$ & 99\% & 99\% & 70\% & 63\% \\
\hline

\end{tabular}
\tablefoot{Table divided into two sections corresponding to different metallicities: $Z = 0.0002$ (top) and $Z = 0.002$ (bottom). In each section, rows represent the initial cluster mass, $\log(M_{\mathrm{cl},0}/\,\Ms)$, and columns correspond to the initial half-mass radius $R_{\mathrm{hm},0}$ in pc.}
\end{table}

Figure \ref{fig:stat_imbh} further illustrates the statistical properties of the IMBH mass distribution as a function of the density for the two metallicities. In both cases, the highest median IMBH mass is found at intermediate core densities, $\rho_{\rm 0,core} = 10^8-10^9\,\Ms\,{\rm pc}^{-3}$, beyond which it decreases as the density further increases. As discussed above, this trend is particularly pronounced at $Z = 0.002$, where the more efficient mass loss due to stellar winds significantly limits the final IMBH mass, capping it at approximately $5\times10^3\,\Ms$. This behavior is not observed in clusters with $M_{\rm cl,0} = 10^5\,\Ms$, for which the median stellar mass increases only mildly with central density at both metallicities. In these systems, the lower masses attained during the runaway phase result in weaker stellar winds, reducing the impact of wind-driven mass loss. Interestingly, for the most compact clusters with $R_{\rm hm,0} = 0.01$ pc and initial masses of $M_{\rm cl,0} = 10^7-10^8\,\Ms$, we again find the formation of massive BHs in a non-negligible fraction of the population. This behavior is related to the interplay between the runaway timescale and our rejuvenation prescription. At such high densities, the VMS accretes the cluster core over a very short timescale, leaving limited time for subsequent rejuvenation. In contrast, in the more massive clusters at intermediate densities, stellar collisions can proceed over a longer period, allowing repeated rejuvenation episodes. Although the mass contributed by individual stars is small, the cumulative effect of successive collisions repeatedly rejuvenate the VMS, extending its effective lifetime and, consequently, the period over which it loses mass through stellar winds. This prolonged wind-driven mass loss ultimately lead to the formation of less massive BHs. We provide a more detailed assessment of the impact of our mass-loss prescriptions in Appendix \ref{app:mass_loss}.

\begin{figure*}
    \centering
    \includegraphics[width=1.\textwidth]{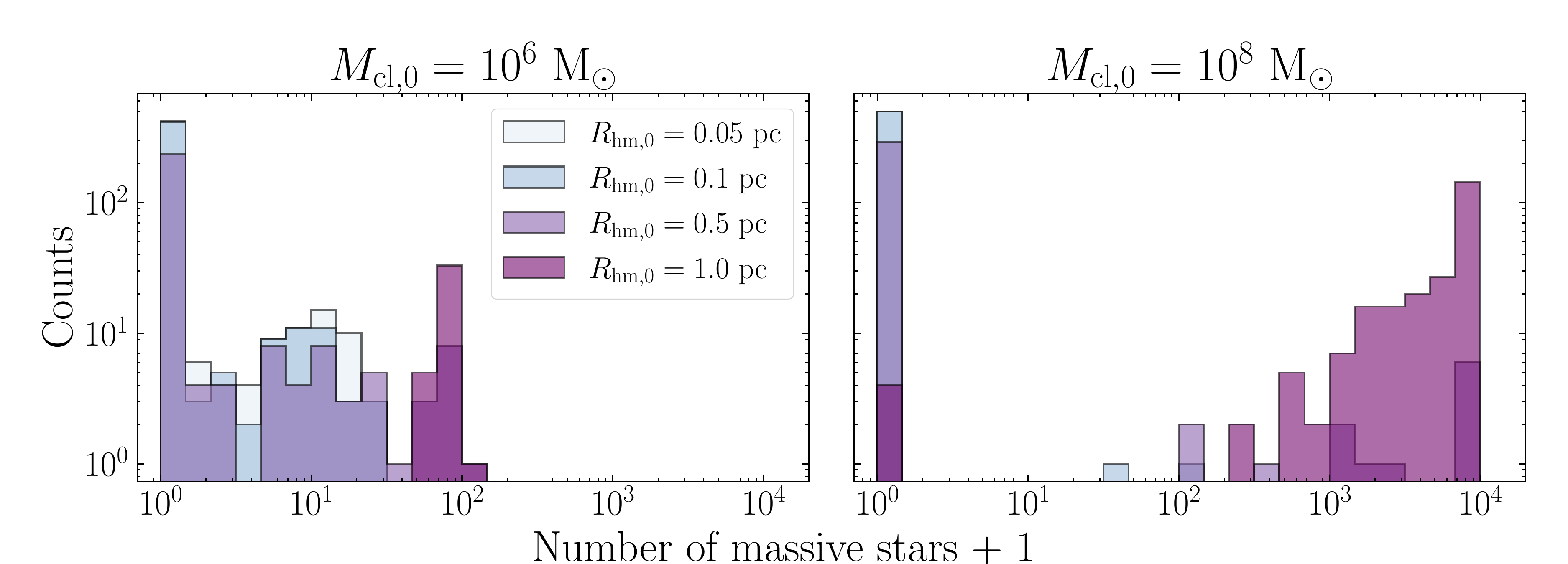}
    \caption{Histograms showing the number of massive stars ($m_* > 20 \,\Ms$) + 1 present in the cluster core given that an IMBH has formed.  From left to right the initial masses of the clusters are $M_{\rm cl} = 10^6\,,\,10^8\,\Ms$, and half-mass radii range from $R_{\rm hm,0} = 0.05$ pc (light blue) to 1 pc (violet).}
    \label{fig:n_massive}
\end{figure*}

In the densest clusters, newly formed BHs are expected to undergo further mass growth through subsequent dynamical interactions and accretion processes. To understand which mechanisms are likely to lead to the mass growth of BHs, it is important to assess how many massive stars are still available in the cluster. Figure \ref{fig:n_massive} shows the distribution of the number of massive stars ($m_* \geq 20\,\Ms$) not accreted by the forming VMS. Less compact clusters generally have a larger number of massive stars, while more compact systems tend to deplete them earlier. In models where zero or only a few massive stars survive in the core, subsequent IMBH growth is expected to proceed mainly through tidal disruption events (TDEs) involving low-mass stars. We discuss these aspects in more detail in the next section.

\section{Discussion}
\label{sec:discussion}
In the following, we discuss the implications of IMBH seed formation for different cluster populations,
and the efficiency of subsequent growth channels.

\begin{figure*}[!ht] 
    \centering
    \includegraphics[width=\textwidth]{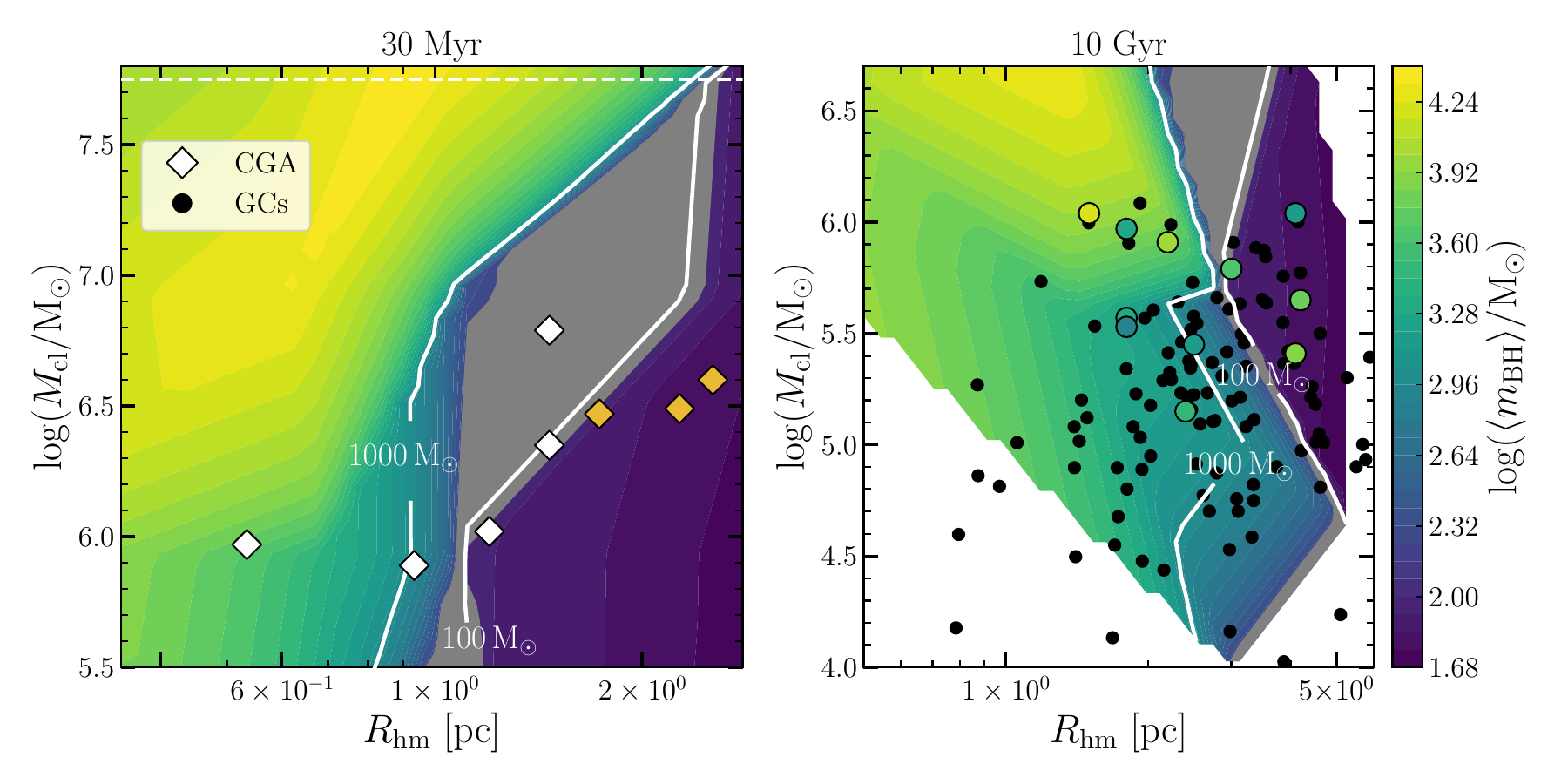}
    \caption{Contour plot in the $M_{\rm cl}-R_{\rm hm}$ plane at 30 Myr (left panel) and 10 Gyr (right panel), showing the mean mass of IMBH seeds obtained from our simulations at $Z = 0.0002$ in 5 Myr. The white (orange) squares mark the masses and half-mass radii (of the counter-images) of the clusters in the Cosmic Gems Arc (CGA) reported by \cite{Messa2026}, while black dots indicate the globular clusters (GCs) from \cite{Harris2015}. Colored dots highlight those GCs for which an estimate of the IMBH mass is available \citep{Lutzgendorf2013,Kiziltan2017}, with the color corresponding to the IMBH mass as indicated by the colorbar. The dashed white line represents the total mass of the Cosmic Gems Arc \citep{Bradley2025}. The grey region highlights the PISN mass gap.}
    \label{fig:gc_gems}
\end{figure*}

To do so, in Figure \ref{fig:gc_gems}, we show the IMBH seed masses predicted by our model after 5 Myr for clusters with masses and radii consistent with those observed in GCs and young massive clusters in CGA, in order to assess whether these systems may have experienced conditions conducive to runaway collisions in the past. To this end, we evolved the mass and half-mass radius of our clusters up to 30 Myr and 10 Gyr. Some of our densest clusters evaporate before reaching 10 Gyr (i.e., those with $\log(M_{\rm cl,0}/\,\Ms) = 5$ and $R_{\rm hm,0} < 0.5$ pc, and $\log(M_{\rm cl,0}/\,\Ms) = 6$ and $R_{\rm hm,0} < 0.1$ pc), and are therefore not included in our analysis at late times.

Aside from CGAs and GCs, we also discuss implications for LRDs, relying on their inferred central densities and structural parameters to evaluate their capability to nurture IMBH seeds, as suggested by other studies \citep[e.g.][]{Pacucci2025}.

\subsection{Implications for Little Red Dots}
\label{sec:LRDs}
Little Red Dots (LRDs) are extremely compact, red, high–surface-brightness sources identified at high and medium redshift \citep{Labbe2023, Kocevski2023,Matthee2024}, whose small sizes and dense stellar components suggest they may host intense central activity. Their origin remains highly uncertain, and several explanations have been proposed, ranging from gas‑enshrouded AGN \citep{Inayoshi2025} to quasi‑stars \citep{Begelman2026}, to self‑gravitating disks accreting onto a supermassive star \citep{Zwick2026}.

\cite{Guia2024} suggested that the inferred stellar masses of LRDs imply extremely high central densities, with a significant tail of the distribution reaching $10^{5}-10^{8}\,\Ms\,\mathrm{pc^{-3}}$, thereby favoring the onset of runaway collisions. As shown in Fig. \ref{fig:stat_mass}, at such densities our models predict the formation of VMSs with masses above $10^{3}\,\Ms$ for clusters with initial masses of $M_{\rm cl,0} = 10^{6},10^{7}\,\Ms$, which in turn produce IMBHs with masses between $\sim 10^3$ and $10^4\,\Ms$. 

However, the typical stellar mass associated with LRDs is $\gtrsim 10^{8}\,\Ms$ \citep{Labbe2023,Akins2025}. The case of $M_{\rm cl,0} = 10^8\,\Ms$ and $R_{\rm hm,0} = 1\,\mathrm{pc}$ is particularly interesting, as the growth of the VMS is largely driven by the intrinsic stochasticity of dynamical interactions. Although the median VMS mass points towards $\sim 200\,\Ms$ (Tab. \ref{tab2}), the distribution is remarkably broad, with a mean mass of a few $\times\,10^3\,\Ms$ (Fig. \ref{fig:contour_coll_vms_growth}, but see also Appendix \ref{app:imbh_seed}) and tails extending beyond $10^4\,\Ms$, naturally resulting in the formation of IMBHs in a fraction of the realizations (Fig. \ref{fig:stat_imbh}). At even higher densities, however, the system enters again a regime of extremely rapid VMS growth. In this regime, the VMSs quickly reach very high maximum masses, after which their evolution becomes dominated by strong stellar-wind mass loss, ultimately leading to the formation of less massive BHs.

Then, our results suggest that, despite the high stellar densities, LRDs can mainly sustain the growth of relatively light VMSs ($\sim10^2-10^3\,\Ms$). Other works focused on the same topics, instead, found that LRD-like systems could lead to VMS up to $10^3-10^4\, \Ms$ (e.g. \citealp{Pacucci2025,Williams2026,Tanikawa2026}). We note that our cluster models inherently assume that cluster contraction is halted by VMS growth, as VMS accretion is halted after the mass in the core has been completely consumed by the VMS itself. Our approach is thus slightly different from the one followed by \cite{Pacucci2025}, which assumes a Fokker-Planck approach to resolve the differential collapse of the innermost layer of a $10^{11}\,\Ms$ Plummer sphere. Interestingly, the approach of \citet{Pacucci2025} leads to central clusters with around $10^6$ stars and stellar densities $\sim 10^7\,\Ms\,\text{pc}^{-3}$, which sustain the VMS growth up to $10^3-10^4\,\Ms$. Under the same assumption of masses and stellar densities, \starfall\ is able to retrieve similar VMS masses, as shown in our Fig. \ref{fig:contour_coll_vms_growth}. 

We also note that, although gas is likely to be present in these environments, we do not consider gas inflows in this work. The interplay between gas accretion and the inward transport of stellar mass may play a key role in the formation of central massive objects (e.g., \citealp{Liempi2026b}, but see also \citealp{Boekholt2018,Tagawa2020,Schleicher2022}).

\subsection{Implications for clusters in the Cosmic Gems Arc}
\label{sec:env_imbh_CGA}

The clusters in the CGA have typical masses of $\sim 10^{6}-10^{7} \,\Ms$ and half-mass radii up to a few parsecs, indicating very dense environments, with ages spanning $\sim 5 - 40$ Myr and metallicities between $Z \sim 0.0004-0.001$ \citep{Adamo2024,Messa2026}.

As already shown in Fig. \ref{fig:contour_coll_vms_growth}, about 60\% of models consistent with clusters in the CGA can host a VMS with mass $> 1000\,\Ms$. The formation of such massive stars in high-redshift systems may be connected to the nitrogen-rich chemical signatures observed in early galaxies, which have been interpreted as evidence of efficient CNO processing in VMSs \citep{Gieles2018,Gieles2025,Charbonnel2023,Williams2026}. In this context, the D.1 ($\log(M_{\rm cl}/\,\Ms) = 5.89$, $R_{\rm hm} = 0.93$ pc) and E.1 ($\log(M_{\rm cl}/\,\Ms) = 5.97$, $R_{\rm hm} = 0.53$ pc) clusters are particularly favorable environments for the formation of IMBH seeds with masses close to or heavier than $1000\,\Ms$ via runaway collisions. Using a Fokker–Planck approach, \cite{Bocchi2026} found that clusters with properties similar to the one inferred for those in the CGA can host IMBH seeds with masses in the range $1600-2700\,\Ms$, consistent with our results. The remaining compatible models fall really in or below the PISN mass gap. This may point to the importance of additional formation channels for IMBHs.  In addition, the CGA is a strongly lensed system \citep{Adamo2024,Bradley2025,Messa2026}. The counter-images represent the lensed projections of the same sources, providing an independent insight on its structural and photometric properties. If we consider the counter-images of these systems (orange squares in Fig. \ref{fig:gc_gems}), these systems do not produce an IMBH seed. However, the C.2+D.2 system is blended and is treated as a single cluster, which may affect its inferred parameters. A more accurate determination of its intrinsic size would help to better constrain its position in parameter space.

In addition, these systems are not expected to fully deplete their massive stellar population. This regime may therefore favor the coexistence of an IMBH and a residual population of massive stars, enabling additional growth through mass transfer or subsequent BBH mergers. In this case, the number of massive stars still available could also favor the formation of additional VMSs through dynamical interactions, which may then collapse into IMBHs, promoting the formation of IMBH–IMBH binaries. Note that upon merger,  the remnant would receive a recoil that, given the typical mass-ratios, hardly exceeds $20 \text{ km s}^{-1}$,  i.e. smaller than the typical escape velocity from such environments (see, e.g., \citealp{Islam2026}). Interestingly, in a recent work, \cite{Torniamenti2026b} showed that clusters with properties similar to those in the CGA can actually form and retain IMBHs with masses exceeding $10^3 \,\Ms$ through hierarchical BH mergers.

\subsection{Implications for globular clusters}
\label{sec:env_imbh_GC}

Observed GCs with half-mass radii $R_{\rm hm} \lesssim 5$ pc and masses $M_{\rm cl} \gtrsim 10^{4.5}\,\Ms$ represent promising candidates for hosting IMBH seeds formed via runaway collisions. Indeed, for a significant fraction of GCs, our models show a $\sim 50\%$ probability of forming VMSs $\gtrsim300\,\Ms$ (Fig. \ref{fig:contour_coll_vms_growth}). These stars have also been proposed as potential contributors to the anti-correlations of nitrogen or helium with respect to oxygen in local globular clusters \citep{Gratton2019,2023Carretta}.  However, the BH masses estimated for most of these systems \citep{Lutzgendorf2013,Kiziltan2017,Haberle2024,BanaresHernandez2025}, and shown in Fig. \ref{fig:gc_gems}, are up to almost one order of magnitude larger than those predicted by our model. This unsurprising discrepancy may be due to the lack of a treatment to model stellar accretion onto the IMBH, and indicates that runaway collisions alone are unlikely to account for the full IMBH mass budget. Considering the IMBH masses in \citet{Lutzgendorf2013} and the ages of the respective host clusters \citep{Catelan2006,Mucciarelli2013,VandenBerg2013,2023Carretta,Gontcharov2026}, we estimated a growth rate in the range $\sim (0.05-0.9)\,\Ms$/Myr. However, several studies (see, e.g., \citealt{Kremer2020,ArcaSedda2023,DragonIII,Torniamenti2024}) have shown that hierarchical BBH mergers in GCs can be suppressed by cluster expansion, thereby limiting mass growth through this channel. In addition, during the runaway phase, massive stars are preferentially merged onto the forming VMS, due to their higher interaction probability in the cluster core. While this process efficiently builds up the VMS mass, at the same time it reduces the number of massive stars available to evolve into BHs and participate in later dynamical interactions (see also Fig. \ref{fig:n_massive}). In the lower-mass clusters considered here ($M_{\rm cl,0} = 10^5$–$10^6\,\Ms$), the reservoir of massive stars in the core is limited to at most a few hundreds. In compact systems ($R_{\rm hm,0} \lesssim 0.5$ pc), this reservoir can be depleted within a few thousand years, thereby limiting the efficiency of subsequent hierarchical BBH mergers. A similar behavior is also found in more massive clusters ($M_{\rm cl,0} = 10^{7}$–$10^{8}\,\Ms$) with small half-mass radii ($R_{\rm hm,0} \lesssim 0.1$ pc), potentially progenitors of the densest and most massive GCs observed today. Although these systems host a larger initial population of massive stars, their extreme densities favor rapid and efficient accretion onto the VMS, again depleting the reservoir of BH progenitors and suppressing the efficiency of hierarchical BBH mergers.

As a consequence, once an IMBH seed has formed, alternative growth channels are expected to become relevant. TDEs involving low-mass stars may dominate the early growth, potentially increasing the BH mass by $\sim 20\%$ within a few Myr (e.g., \citealp{Rantala2026}, but see also \citealp{Stone2017,Rizzuto2023}). On longer timescales, comparable to the ages of observed GCs, TDEs involving low-mass compact remnants (e.g. white dwarfs) may contribute both to mass growth and to observable high-energy transients \citep{Giersz2015,Tanikawa2022,DragonIII}. IMBH–white dwarf systems have been proposed as potential progenitors of long gamma-ray bursts \citep{Oganesyan2025,Yuan2026}, making them especially relevant in the context of multi-messenger astrophysics and IMBH mass growth.

On the other hand, when considering massive clusters with larger half-mass radii ($R_{\rm hm,0} \sim 0.5$ – $1$ pc) that form an IMBH, several hundred or even thousands of massive stars may remain. Given the large number of stars available in the core, further dynamical interactions can lead to the formation of additional VMSs, which may in turn collapse into massive BHs or IMBHs. If these BHs are retained, such clusters provide favorable environments for the formation of IMBH–BH or IMBH–IMBH binaries. In such a case, relativistic recoils could become comparable to or larger than the cluster escape velocity, lowering the retention probability.

\subsection{Uncertainties and limitations}

The present implementation of \starfall\ relies on some simplifying assumptions that may affect the generality of our results. 

First, the adopted stellar evolution treatment neglects the effects of binary stellar evolution, such as mass transfer and common envelope phases. Although these processes can significantly alter the binary properties, they typically occur on timescales longer than those explored in our simulations, and are therefore expected to have a limited impact on our conclusions. 

When two stars merge or collide, the radius of the remnant is computed as presented in Sect. \ref{sec:stellar_evolution}, where $m_*$  is the sum of the masses of the stars involved in the interaction, corrected for mass loss. This approximation is partially supported by hydrodynamical simulations of stellar collisions \citep{Glebbeek2009}; however, the actual radius may depend on the collision geometry, the internal structure of the progenitors, their masses, and how rapidly the star relaxes to its compact state. Indeed, after a collision the remnant could be initially expanded, and Fig.~\ref{fig:vms_mass_growth} shows that the primary star can experience multiple collisions within just a few thousand years in dense environments. This timescale may be shorter than the time required for the star to contract before undergoing a new interaction (see, e.g., \citealp{Costa2022}). Indeed, when multiple collisions occur out of equilibrium, the resulting VMS remnant is likely to be puffed up, increasing its cross-section for contact with other stars \citep{Dale2006}. This may lead to an underestimation of the interaction rate. However, a larger cross-section does not necessarily translate into a higher merger rate, as intruding stars may pass through the puffy envelope without being captured or merging with the VMS. This may therefore lead to an underestimation of the interaction timescales. On the other hand, \cite{Vergara2026} recently showed that using the cross‑section argument to compute interaction timescales can result in an overestimation when a massive VMS interacts with a few stars left in the core. At each timestep, nevertheless, we limit the number of mergers and collisions to the number of objects within the VMS's radius of influence. In addition, the prescription for mass loss adopted here for stellar collisions and mergers may break down in the densest and most massive clusters we considered. In such environments, the velocity dispersion can approach the stellar escape velocity, reducing the amount of mass that remains bound to the remnant.

As mentioned in Sect. \ref{sec:stellar_evolution}, we determine the final fate of a star based on its pre-SN mass, although the PISN regime is primarily determined by the He-core properties. Most interactions in the collision chains shown in Fig. \ref{fig:vms_mass_growth} involve two main-sequence stars ($\sim99\%\,,\,\sim60\%$ when $R_{\rm hm} = 0.05,0.1$ pc respectively), for which assuming a merger remnant structure similar to that of the corresponding rejuvenated star is a reasonable approximation. A different situation can arise when a post-main-sequence primary collides with a main-sequence star. In the densest clusters ($R_{\rm hm}=0.05\,,\,0.1$ pc), this may happen when the primary star is already above the PISN mass gap ($m_* \sim \text{ few}\times10^3\,\Ms$). In the less dense cluster ($R_{\rm hm}=1$ pc), the post-main-sequence star may undergo only one or two collisions at lower masses and subsequently enter the PISN mass gap. To first order, the relevant quantity for determining the final fate in this case is the mass of the primary's He core. Given our initial conditions, the primary has a core mass that leads to a PISN, allowing us to reasonably assume that the remnant will undergo the same fate. Interactions between two post-main-sequence stars with a collision remnant entering the pair-instability regime, which would require a more detailed treatment of the remnant's internal structure, are rare ($\lesssim\,0.1\%$) and will be investigated in future work.

From a dynamical point of view, \starfall\ includes binary-single interactions and direct stellar collisions, neglecting binary-binary encounters that may become dominant in environments with a high binary fraction (e.g., \citealt{Mikkola1983,Zevin2019,ArcaSedda2021a}). In addition, the code does not account for the formation and evolution of hierarchical triple systems arising from dynamical interactions. Such systems can drive large-amplitude eccentricity oscillations in the inner binary through the Kozai-Lidov mechanism \citep{Kozai1962,Lidov1962}, potentially leading to stellar collisions. We note, however, that for dense systems stable triples are very improbable \citep{Leigh2013}.

Stellar masses are sampled assuming a Kroupa IMF \citep{Kroupa2001} up to $150\,\Ms$. While this choice is reasonable for Population I and II stars, it remains uncertain for Population III, for which both the shape and the limits of the mass function, as well as the properties of binary systems, are still poorly constrained \citep{Bromm2004,Glover2013,Liu2021,Prole2022}. 

Finally, we note that our results should be regarded as conservative. Our simulations focus on stars initially located in the cluster core and neglect contributions from the outer regions. However, strong mass segregation episodes and primordial mass segregation can funnel more massive stars toward the center \citep{Rantala2024}, potentially leading to a more massive VMS and increasing the number of BHs in the core, with implications for the subsequent growth of the central BH.


\section{Summary and conclusions}
\label{sec:summary}

In this work, we investigated the formation of VMSs and IMBH seeds through runaway collisions and stellar mergers in dense stellar clusters. 

We developed \starfall, a semi-analytic tool designed to follow the evolution of a massive stellar binary in the cluster core, accounting for binary-single interactions, stellar mergers, and collisions, as well as the dynamical evolution of the host cluster, with particular focus on core collapse (Sect. \ref{sec:methods}).

We showed that in star clusters with masses in the range $M_{\rm cl,0} = 10^6-10^8\,\Ms$ and half-mass radii $R_{\rm hm,0} < 1$ pc, stellar collisions are more likely to occur and can lead to the formation of VMSs with masses $m_* \gtrsim 10^3\,\Ms$ (Figs. \ref{fig:vms_mass_growth}, \ref{fig:contour_coll_vms_growth}). The maximum VMS masses are strongly correlated with the initial mass and density of the host cluster (Fig. \ref{fig:stat_mass}). In addition, the cluster structure, its dynamical evolution, and the relevant stellar evolutionary timescales play a key role in determining the probability of VMS formation. Even in clusters with core-collapse timescales longer than 5 Myr, the probability of forming a VMS in our model remains significant (between 15\% and 50\%), provided that the cluster is massive enough ($M_{\rm cl} \gtrsim 10^7\,\Ms$).

We compared the maximum stellar masses reached at the end of the collisional chains with results from $N$-body simulations presented by \citealt{ArcaSedda2023,Vergara2026,Rantala2026}; Askar et al. in prep.; Giersz et al., in prep. (Fig. \ref{fig:stat_mass}). We find overall good agreement, with \starfall\ also probing the high-density tail of the parameter space, which is not accessible to $N$-body simulations.

We also showed that cluster models with properties similar to those observed in the CGA \citep{Adamo2024,Bradley2025,Messa2026} can host VMSs with $m_* \geq 10^3\,\Ms$ in $\sim 60\%$ of the cases (Fig.~\ref{fig:gc_gems}), indicating that such environments are particularly favorable for the formation of both VMSs and IMBHs. However, uncertainties in the cluster sizes can significantly affect their position in parameter space, potentially shifting the predicted IMBH seed masses toward, or even below, the PISN mass gap.

Globular clusters (GCs) also represent favorable environments for forming VMSs with $m_* \geq 300 \,\Ms$ and IMBH seeds (Fig.~\ref{fig:gc_gems}; \citealt{Lutzgendorf2013,Harris2015,Kiziltan2017}). In dense, low-mass GCs, where the population of massive stars has already been largely depleted (Fig. \ref{fig:n_massive}), alternative channels to BH-BH mergers may become dominant for subsequent mass growth of the IMBH. In particular, TDEs involving low-mass stars \citep{Stone2017,Rizzuto2023,Rantala2026} are expected to play a key role in driving further mass growth.

In contrast, although LRDs may have high stellar masses and densities \citep{Guia2024,Akins2025}, their structural properties, if interpreted as stellar clusters, do not always favor the onset of long stellar collision chains. For core densities between $10^5$ and $10^8\,\Ms\,\text{pc}^{-3}$ and cluster masses of $10^6$–$10^7\,\Ms$, IMBHs with masses in the range $10^3$–$10^4\,\Ms$ can form. For more massive clusters ($M_{\rm cl}=10^8\,\Ms$), the longer core-collapse timescale results in a lower median IMBH mass. Although the distribution is very broad, the probability of forming BHs with masses above $10^3\,\Ms$ remains low (Sect. \ref{sec:LRDs}). This suggests that high density alone is not sufficient to trigger collision chains. Resolving their internal structure will play a key role in determining their dynamical evolution. However, \cite{Pacucci2025} suggest that, in these high density systems, mass segregation happens on very short timescales favoring the onset of runaway collisions and the formation of more massive VMSs. Nevertheless, when considering the same initial conditions ($M_{\rm cl, 0} \sim 10^6\,\Ms$, $R_{\rm hm,0} \sim 0.1$ pc), we find consistent results.

We further deepened the impact of metallicity and stellar wind mass loss on the mass of IMBH seeds (Figs. \ref{fig:histo_vms_imbh},\ref{fig:stat_imbh}; but see also Appendix \ref{app:mass_loss}). We found that, although denser clusters tend to produce more massive VMSs, they do not necessarily produce the most massive BHs, with clusters at intermediate densities ($\rho_{\rm 0,core} \sim 10^7-10^9 \,\Ms\,{\rm pc}^{-3}$) producing IMBHs with masses higher than those produced in denser environments. This introduces a dependence on both the formation timescale of the VMS and its subsequent evolution (see also Fig. \ref{fig:vms_mass_growth}).

While our work identifies the regimes in which VMSs with masses above $10^3\,\Ms$ can form, it also highlights the critical role of the subsequent stellar evolution once these objects reach their maximum mass. Future studies that accurately model the evolution of collision products, together with improved prescriptions for stellar winds, will be essential to further refine these results.


\begin{acknowledgements}

We thank the anonymous referee for their careful reading of the manuscript and their valuable comments, which improved the clarity and quality of this work. The authors gratefully acknowledge Antti Rantala, Michela Mapelli, Filippo Santoliquido, Gor Oganesyan and Dominik Schleicher for the helpful comments and suggestions. MAS acknowledges funding from the European Union’s Horizon 2020 research and innovation programme under the Marie Skłodowska-Curie grant agreement No.~101025436 (project GRACE-BH) and from the MERAC Foundation through the 2023 MERAC prize. MAS and MB acknowledge the ACME project which has received funding from the European Union's Horizon Europe Research and Innovation programme under Grant Agreement No.~101131928. GI is supported by a fellowship grant from la Caixa Foundation (ID 100010434). The fellowship code is LCF/BQ/PI24/12040020. PB thank the support from the special program of the Polish Academy of Sciences and the U.S. National Academy of Sciences under the Long-term program to support Ukrainian research teams, grant No.~PAN.BFB.S.BWZ.329.022.2023. AA acknowledges that this research was funded in part by National Science Centre (NCN), Poland, grant No. 2024/55/D/ST9/02585. For the purpose of Open Access, the author has applied a CC BY public copyright licence to any Author Accepted Manuscript (AAM) version arising from this submission. PB also appreciate the Polish high-performance computing infrastructure PLGrid (HPC Centre: ACK Cyfronet AGH -- docs.hpc.cyfronet.pl) for providing Helios computer facilities and support within computational grant No.~PLG/2026/019243. GC acknowledges financial support from the European Union–Next Generation EU, Mission 4, Component 2, CUP: C93C24004920006, project ‘FIRES’. MG was supported by the Polish National Science Center (NCN) through the grant 2021/41/B/ST9/01191. ST acknowledges financial support from the Alexander von Humboldt Foundation for the Humboldt Research Fellowship. MCV acknowledges financial support from the Center for Astrophysics and Associated Technologies (CATA; FB210003). We used \bpop\ (\url{https://github.com/marcasedda/BPOP}) to model star cluster evolution \citep{ArcaSedda2020,ArcaSedda2023,ArcaSedda2026}. We used \sevn\ (\url{https://gitlab.com/sevncodes/sevn}; \citealt{Spera2019,Iorio2023}) to interpolate \textsc{Parsec} stellar tracks \citep{Costa2025}. This research made use of \textsc{NumPy} \citep{Harris20}, \textsc{SciPy} \citep{SciPy2020}, \textsc{Pandas} \citep{Pandas2024}. For the plots we used \textsc{Matplotlib} \citep{Hunter2007}.
\end{acknowledgements}

\section*{Data availability}

The \starfall\ code will be publicly available soon. In the meantime, the code can be made available upon request. Interested users are encouraged to contact the authors.

%

\bibliographystyle{aa}
\bibliography{ref}


\begin{appendix}




\nolinenumbers

\section{Star cluster model}
\label{app:bpop}
In \starfall, star clusters are modeled following the prescriptions in the \bpop\ code \citep{ArcaSedda2020,ArcaSedda2023,ArcaSedda2026}. 

Starting from a cluster mass $M_{\rm cl,0}$, the evolution of the bound mass is described by:
\begin{equation}
M_{\rm cl}(t) = M_{\rm cl,0}\left(1+\frac{t}{t_{\rm st}}\right)^{k} \left(1-\frac{t}{f(M_{\rm cl,0})t_{\rm rlx}}\right),
\end{equation}
where $t_{\rm st} = 10\,\mathrm{Myr}$, $k = -0.1$, and $t_{\rm rlx}$ is the relaxation time. The parameter:
\begin{equation}
f(M_{\rm cl,0}) \equiv \alpha_{\rm bnd} \left( \frac{M_{\rm cl,0}}{10^6\,\Ms} \right)^{\beta_{\rm bnd}},
\end{equation}
regulates the onset of relaxation-driven mass loss. We adopted $\alpha_{\rm bnd}=1000$ and $\beta_{\rm bnd} = 0.6$. These parameters are calibrated to reproduce globular cluster evolution reported by \cite{Gieles2023}.

Similarly, given a half-mass radius $R_{\rm hm,0}$, its time evolution follows:
\begin{equation}
R_{\rm hm}(t) = R_{\rm hm,0} \left(1 + \frac{t}{t_0} \right)^{\alpha_{\rm hm}},
\end{equation}
where $t_0 = 5\,t_{\rm cc}\left[1 + \left(R_{\rm hm,0}/1\mathrm{pc}\right)\right]$, $t_{\rm cc}$ is the core-collapse time, and $\alpha_{\rm hm}$ is randomly sampled at each time step from a uniform distribution between $0.15 - 0.25$. This prescription reproduces the half-mass radius evolution observed in the \dragon\ simulations \citep{DragonII,DragonI,DragonIII}.

The core properties are modeled through semi-analytical relations linking the core mass and core radius to $M_{\rm cl, 0}$:
\begin{equation}
M_{\rm core} = \alpha_{\rm core} M_{\rm cl,0} \left(\frac{R_{\rm core}}{0.15\mathrm{pc}}\right)^{\beta_{\rm core}},
\end{equation}
where $\alpha_{\rm core} = 0.03$. The exponent $\beta_{\rm core}$ depends on the evolutionary stage of the cluster and is set to $\beta_{\rm core} = 0.4$ for $t < 2t_{\rm cc}$, and to $\beta_{\rm core} = 1/1.3$ at later times.

The ratio between the core radius and the half-mass radius evolves according to:
\begin{equation} \frac{R_{\rm core}}{R_{\rm hm}} = \begin{cases} 
{\rm max}\left(R_{\rm min}, 0.2 \left(1-\displaystyle \frac{t}{2t_{\rm cc}}\right)^{0.53}\right) & t \leq 2t_{\rm cc},\\ 
R_{\rm min} + 0.2\left(\displaystyle \frac{t}{2t_{\rm cc}} - 1 \right)^{1/4} & 2t_{\rm cc} < t \leq 5 t_{\rm cc},\\
R_{\rm min} - 0.8 + \left(\displaystyle\frac{t}{2t_{\rm cc}}-1\right)^{0.05} & t > 5t_{\rm cc}, 
\end{cases} \end{equation}
where $R_{\rm min} \propto \left(m_{\rm mean}/M_{\rm cl}\right)^{1.3}$, and $m_{\rm mean}$ is the mean stellar mass \citep{Binney2008}.

\section{Binary-single interactions}
\label{app:bs_int}

Binary-single interaction modeling is based on \cite{Heggie1975,Hut1983,Heggie1993,Sigurdsson1993}.

The outcome of a binary–single encounter is set by the ratio of the transferred energy:
\begin{equation}
\Delta E_{\rm bs} \sim \left(\frac{r_{\rm p,e}}{a}\right)^{3/2}
\end{equation}
to the perturber kinetic energy in the three-body center-of-mass frame:
\begin{equation}
\Delta E_{\rm c} = \frac{m_{\rm bin} m_{\rm p}}{2(m_{\rm bin}+m_{\rm p})} v_\infty^2,
\end{equation}
where $a$ is the binary semi-major axis, $r_{ \rm p,e}$ the minimum separation between the binary and the incoming perturber, $m_{\rm bin} = m_1 + m_2$, $m_{\rm p}$ the perturber mass, and $v_\infty$ its velocity at infinity \citep{Heggie1975,Sigurdsson1993}.
If the transferred energy is small ($\Delta E_{\rm bs} < \Delta E_{\rm c}$), the binary outcome is set by its binding energy $E_{\rm bin}$. Specifically, if $\Delta E_{\rm bs} < E_{\rm bin}$, the binary either hardens or softens depending on the local velocity dispersion. Otherwise, the binary undergoes an exchange, replacing its least massive component if the perturber is heavier than the binary or its individual members.

If the binary preserves its components, the perturber will recede to infinity with velocity $v_{\rm p} < v_\infty$. If not, the former binary component with mass $m_{\rm i}$ will escape with velocity 
\begin{equation*}
    \displaystyle\frac{v_{\rm p}}{v_\infty} < \left(\frac{m_{\rm p}}{m_{\rm i}}\right)^{1/2}.
\end{equation*}

High-energy encounters ($\Delta E_{\rm bs} > \Delta E_{\rm c}$) are treated statistically by comparing $v_\infty$ to the critical velocity for ionization $v_{ \rm c}$ \citep{Hut1983}:
\begin{equation}
v_{\rm c}^2 = \frac{G m_1 m_2 (m_{\rm bin}+m_{\rm p})}{a_{\rm bin} m_{\rm bin} m_{\rm p}}.
\end{equation}
For slow interactions ($v_\infty < v_{\rm c}$), the three bodies undergo a resonant encounter, typically resulting in an exchange of a binary component if the perturber is sufficiently massive or the energy transfer is large; otherwise, the lighter component is ejected. In fast interactions ($v_\infty > v_{\rm c}$), the binary either swaps a component if $\Delta E_{\rm bs} < E_{\rm bin}$ or is ionized if $\Delta E_{\rm bs} > E_{\rm bin}$.

The scattering is assumed to occur at $v_\infty = \sigma$, equal to the cluster velocity dispersion, which is a time-varying function that depends on the cluster mass and radius. 

The interaction is assumed to be hyperbolic and fully characterized by the impact parameter and pericenter distance. The squared impact parameter, $b^2$, is drawn from a flat distribution between 0 and $b_{\rm max}^2$, where $b_{\rm max}$ is defined as:
\begin{equation}
    b_{\rm max} = \left[\frac{Cv_{\rm c}}{v_\infty} + 0.6\, (1+e_{\rm bin})\right]\,
    a_{\rm bin},
\end{equation}
with $C = 3$ \citep[see e.g.][and reference therein]{Sigurdsson1993}. The pericenter, instead, is given by:
\begin{equation}
    r_{\rm p} = \frac{G\,(m_{\rm bin}+m_{\rm p})}{v_\infty^2}\left \{\left[1+b^2\left(\frac{v_\infty^2}{G\,(m_{\rm bin}+m_{\rm p})}\right)^2\right]^{1/2} -1\right \},
\end{equation}
and the energy transferred in the interaction is: 
\begin{equation}
    \label{eq:delta_E}
    \Delta E_{\rm bs} = -A\left(\frac{r_{\rm p}}{a_{\rm bin}}\right)^{3/2}\exp\left[-B\left(r_{\rm p}/a_{\rm bin}\right)^3\right].
\end{equation}
The quantities $A$ and $B$ are functions of the binary orbital properties and scattering parameters \citep[see Eqs. 2.12-2.13 in][]{Sigurdsson1993}. However, Eq.~\ref{eq:delta_E} is valid only in the limit $r_{\rm p}/a_{\rm bin} \gg 1$. Since, by construction, we consider hard binaries (see Sect.~\ref{sec:dynamics}), for encounters that do not satisfy this condition, we instead sample the fractional energy change, $x \equiv \Delta E/E_{\rm bin}$, from the distribution derived by \cite{Heggie1975}, who showed that the rate of close encounters producing a fractional change $x$ in the binary binding energy scales as $(1+x)^{-9/2}$.

If the recoil velocity imparted on the binary after the interaction exceeds the cluster escape velocity, $v_{\rm esc} \simeq (2GM_{\rm cl}/R_{\rm hm})^{1/2}$, the binary is ejected from the cluster.

\section{Collision outcomes}
\label{app:contact}
\begin{figure}
    \centering 
    \includegraphics[width=1.\columnwidth]{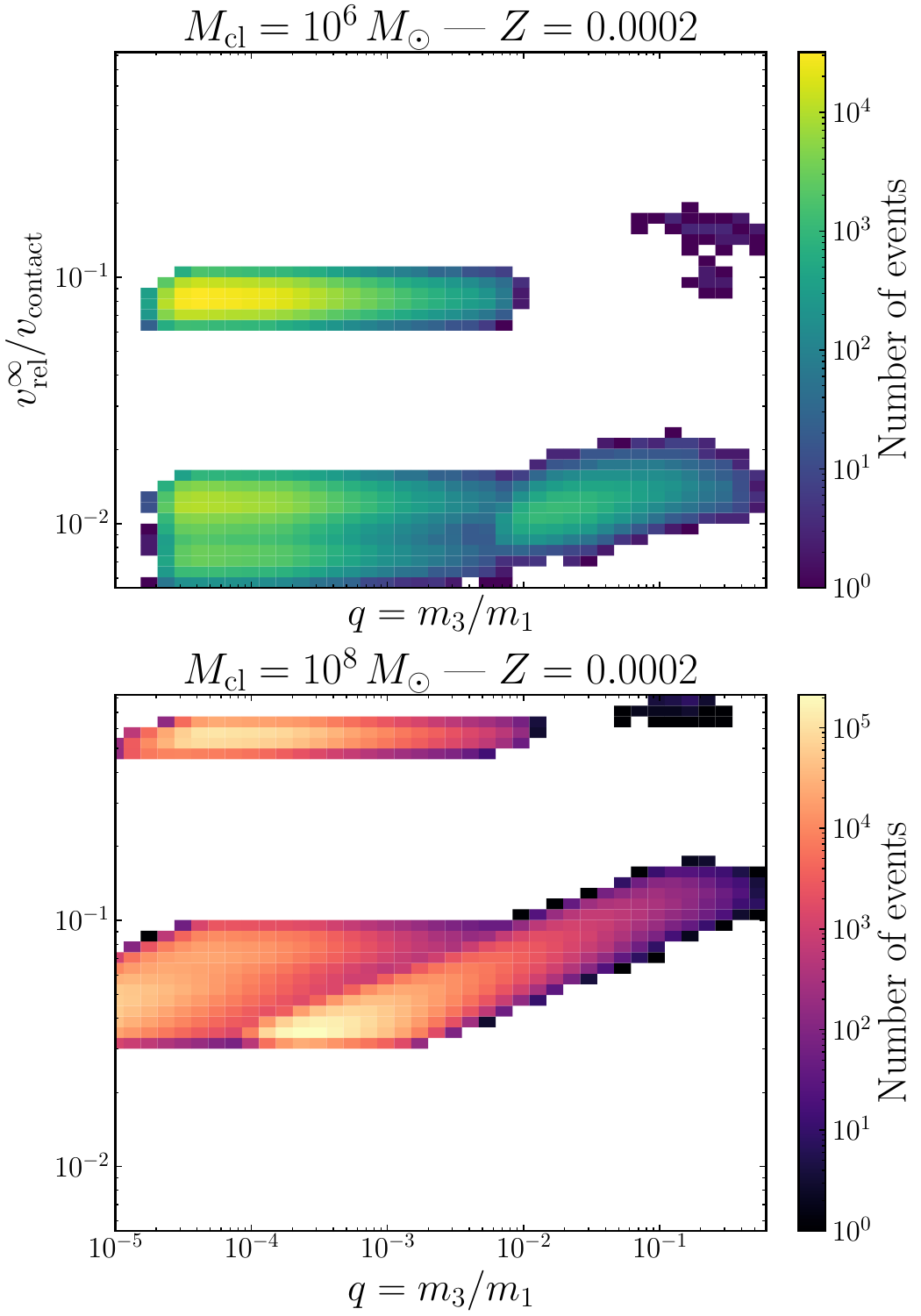}
    \caption{2D histogram showing the mass ratio on the $x$-axis and the ratio $v_{\rm rel}^{\infty}/v_{\rm contact}$ on the $y$-axis for all collisions undergone by the systems considered, for clusters with $M_{\rm cl,0} = 10^6\,\Ms$ (top panel) and $10^8\,\Ms$ (bottom panel) and $R_{\rm hm} = 0.5~\mathrm{pc}$. The color maps indicate the number of events in each bin.} 
    \label{fig:contact} 
\end{figure}
In this Appendix we examine in more detail the assumption that a contact between two stars results in a merger. We note that, when computing the collision timescale, $t_{\rm coll}$, we always assume the pericenter distance, $r_{\rm p}$, to be equal to the sum of the radii of the two stars, i.e. $r_{\rm p}/(R_1+R_2) = 1$, where $R_i$ are the radius of the two star (see Sect. \ref{sec:dynamics}). Then, following \cite{Freitag2005}, we define:
\begin{equation}
    v_* = \sqrt{\frac{2\,G\,(m_1+m_2)}{R_1+R_2}}\,,
    \label{eq:v*}
\end{equation}
\begin{equation}
    v_{\rm contact} = \sqrt{(v_{\rm rel}^{\infty})^2\,+\,v_*^2}\,,
\end{equation}
where $m_{i}$ are the masses of the two stars, and $v_{\rm rel}^{\infty}\,\sim\,\sigma$, with $\sigma$ the velocity dispersion of the cluster. In Fig. \ref{fig:contact} we show the distribution of $v_{\rm rel}^{\infty}/v_{\rm contact}$ as a function of the mass ratio for all collisions recorded in the simulated systems, for two representative cluster masses ($M_{\rm cl,0} = 10^6$ and $10^8\,\Ms$) with $R_{\rm hm,0} = 0.5\,\mathrm{pc}$. We select these two cases because for $M_{\rm cl,0} = 10^6\,\Ms$ the bulk of collisions is expected to occur during core collapse, while the more massive counterpart allows us to assess how the collision properties vary with cluster mass.

The vast majority of events concentrate at $v_{\rm rel}^{\infty}/v_{\rm contact} \lesssim 0.2$, indicating that the relative velocity at infinity is small compared to the contact velocity and that the encounters are gravitationally focused. Furthermore, after the merger of the primordial binary (the necessary first step to initiate the collision chain; see also Sect. \ref{sec:results}), the mass ratio rapidly decreases, driving the system into a regime where, following \cite{Freitag2005}, a merger outcome is well justified. This is particularly relevant for $M_{\rm cl,0} = 10^8\,\Ms$, where a non-negligible fraction of collisions occurs at $v_{\rm rel}^{\infty}/v_{\rm contact} \sim 0.5 - 0.6$; however, these events have mass ratios below $0.05$, which still supports the merger assumption. Only a few isolated events fall in a region of parameter space where the distinction between a fly-by and a merger becomes ambiguous.

Finally, we note that, with respect to \cite{Freitag2005}, we do not resolve the stellar structure, and in Eq. \ref{eq:v*} we therefore adopt the total stellar radius rather than the half-mass radius, which leads to an overestimation of $v_{\rm rel}^{\infty}/v_{\rm contact}$. In addition, as discussed in Sect. \ref{sec:cluster_model}, core collapse is intrinsically halted by the presence of a VMS (or a binary system), preventing the velocity dispersion from reaching very high values. Moreover, after each collision the VMS radius is recomputed according to Eq.~\ref{eq:r_zams}, yielding a relatively compact remnant. Taken together, these considerations justify treating contact events as mergers.

\FloatBarrier

\section{Mass distribution of very massive stars and intermediate-mass black hole seeds} \label{app:imbh_seed}
\begin{figure*} 
    \centering 
    \includegraphics[width=0.9\textwidth]{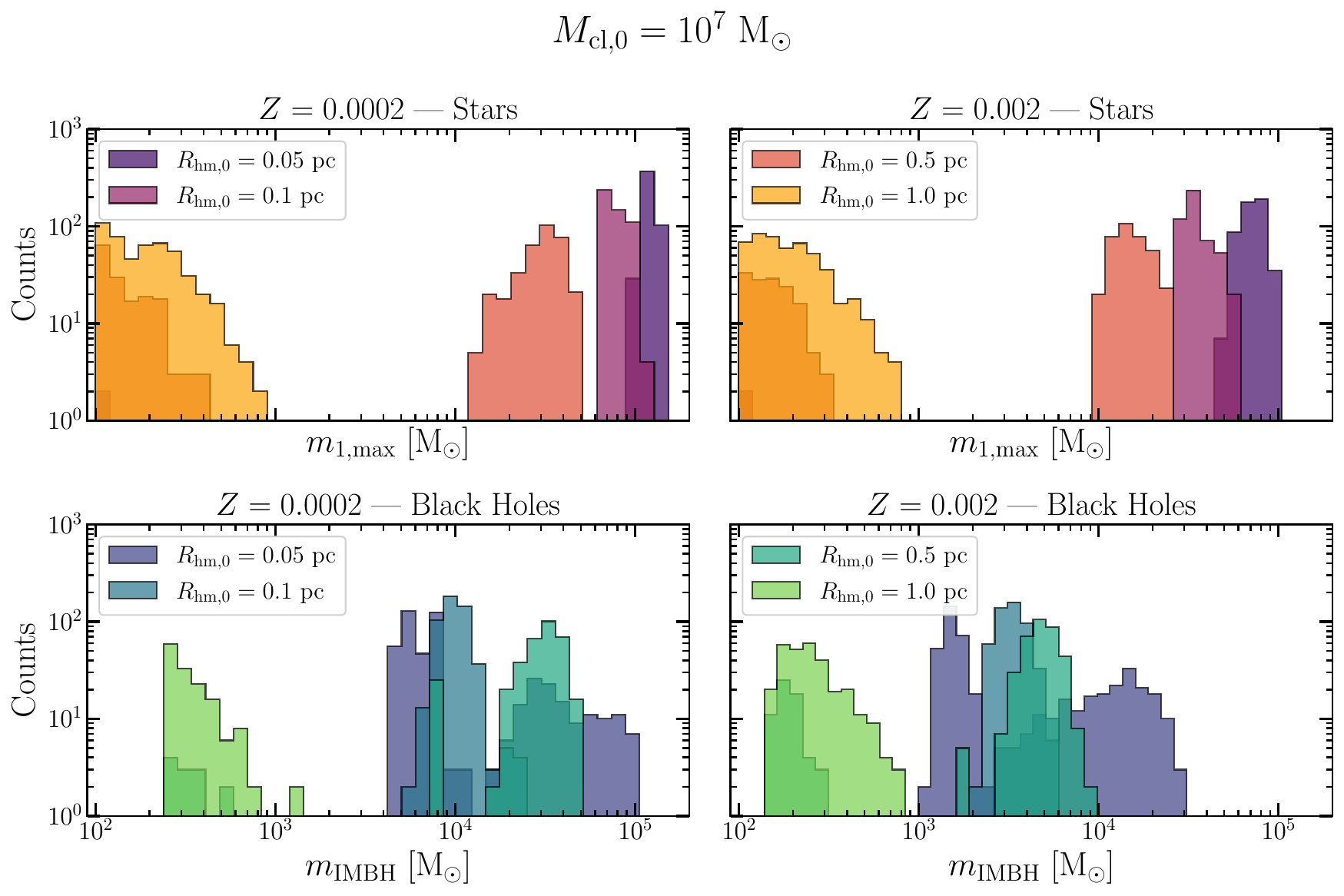}
    \caption{Same as Fig. \ref{fig:histo_vms_imbh}, but for $M_{\rm cl} = 10^7 \,\Ms$.} 
    \label{fig:histo_vms_imbh_7} 
\end{figure*}

\begin{figure*} 
    \centering 
    \includegraphics[width=0.9\textwidth]{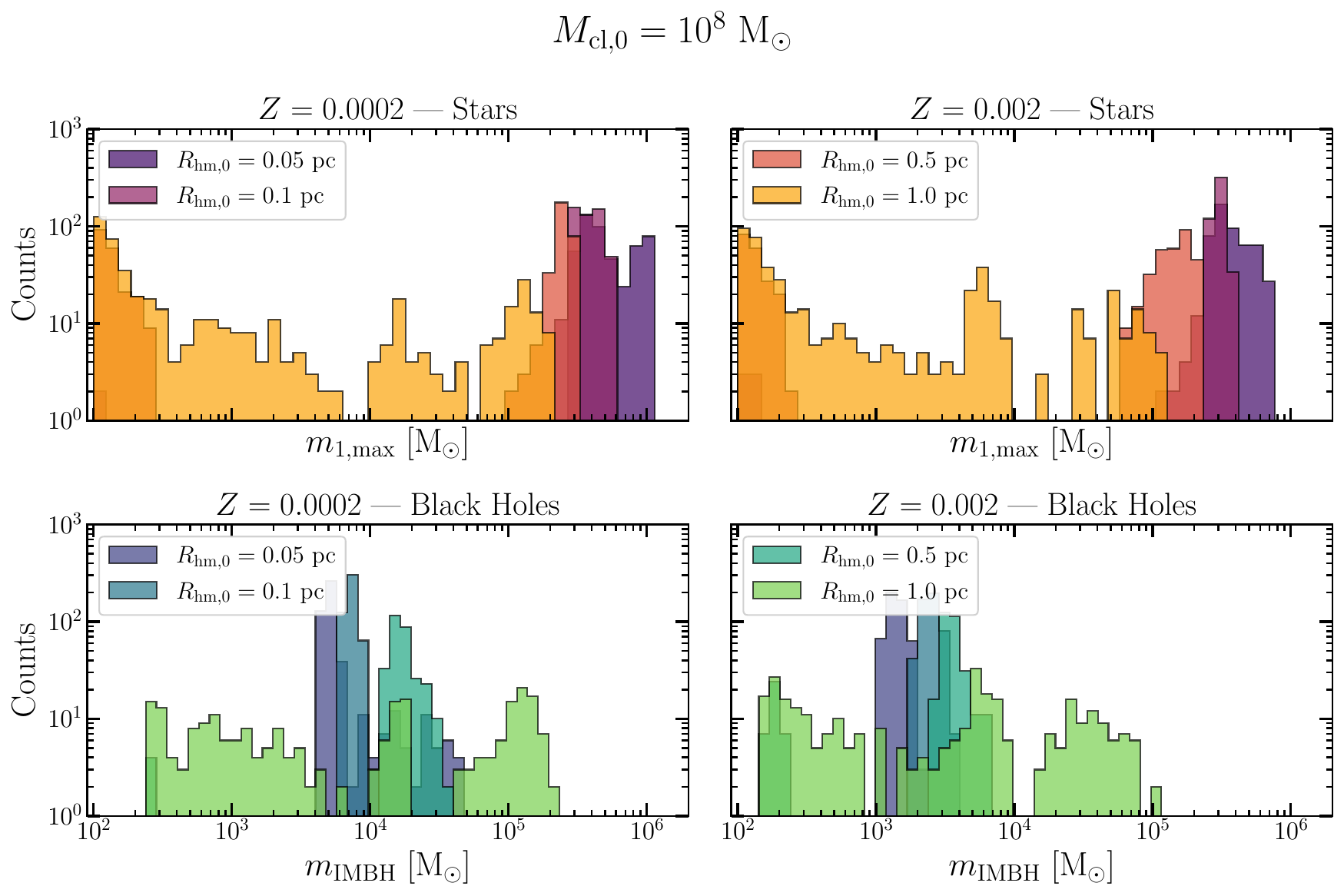} 
    \caption{Same as Fig. \ref{fig:histo_vms_imbh}, but for $M_{\rm cl} = 10^8 \,\Ms$.} 
    \label{fig:histo_vms_imbh_8} 
    \end{figure*}

Figures \ref{fig:histo_vms_imbh_7} and \ref{fig:histo_vms_imbh_8} show the distributions of the maximum stellar mass and IMBH seed mass for clusters with $M_{\rm cl,0} = 10^7$-$10^8 \,\Ms$. The overall trends are similar to those found for $M_{\rm cl,0} = 10^6 \,\Ms$ (Fig. \ref{fig:histo_vms_imbh}), but shifted toward higher masses.

For $M_{\rm cl,0} = 10^7 \,\Ms$ ($10^8 \,\Ms$) and $R_{\rm hm,0} = 0.05,0.1$ pc, the stellar mass distribution exhibits a peak at $\sim 10^5\,\Ms$ ($\sim 6\times10^5\,\Ms$), nearly independent of metallicity ($Z = 0.0002,0.002$). At $R_{\rm hm,0} = 0.5$ pc, nearly half of the realizations produce VMSs with masses exceeding $10^4\,\Ms$ for both cluster masses. At $R_{\rm hm,0} = 1$ pc, however, the two cluster masses exhibit markedly different distributions. For $M_{\rm cl,0} = 10^7,\Ms$, the VMS mass distribution is entirely confined between $10^2$ and $10^3\,\Ms$. At the higher cluster mass, instead, the distribution becomes substantially broader, developing a prominent high-mass tail that extends beyond $10^4\,\Ms$.

For clusters with initial mass $M_{\rm cl,0} = 10^7\,\Ms$, the most massive IMBHs reach $10^5\,\Ms$ for $R_{\rm hm,0} = 0.05$ pc and $\sim5\times10^4\,\Ms$ for  $R_{\rm hm,0} = 0.5$ pc, with higher metallicity reducing BH masses by nearly an order of magnitude. For more massive clusters, instead, $M_{\rm cl,0} = 10^8\,\Ms$, the most massive IMBHs form at $R_{\rm hm,0} = 1$ pc, although only in a few realizations. In this case, he densest clusters do not produce BHs more massive than $\sim 10^4\,\Ms$, regardless of metallicity.

\begin{figure*} 
    \centering 
    \includegraphics[width=0.9\textwidth]{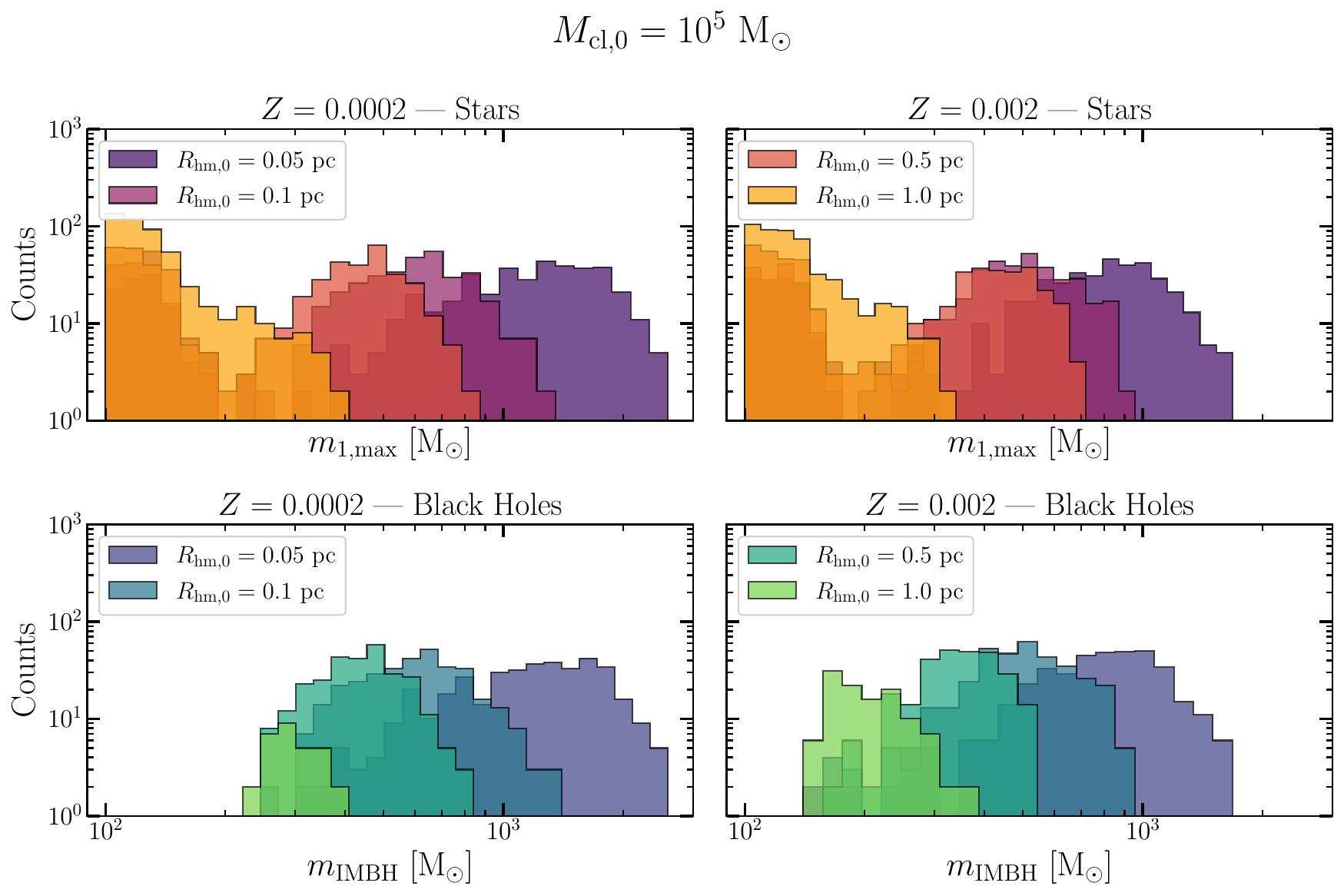}
    \caption{Same as Fig. \ref{fig:histo_vms_imbh}, but for $M_{\rm cl} = 10^5 \,\Ms$.} 
    \label{fig:histo_vms_imbh_5} 
\end{figure*}

For lower-mass clusters ($M_{\rm cl,0} = 10^5 \,\Ms$, Fig. \ref{fig:histo_vms_imbh_5}), this bimodal behavior is not observed. The stellar mass distribution peaks at $\sim (5-7)\times10^2\,\Ms$ for $R_{\rm hm,0} = 0.1$ and $0.5$ pc, whereas it extends to higher masses, reaching $\sim 2\times10^3\,\Ms$, for $R_{\rm hm,0} = 0.05$ pc. Accordingly, the IMBH seed mass can range from $\sim 500\,\Ms$ to $\sim 2000\,\Ms$, depending on the initial cluster density. For $R_{\rm hm,0} = 1$ pc, the smoother dynamical evolution largely suppresses runaway collisions, with only a few exceptions.

\FloatBarrier

\section{Impact of stellar evolution prescriptions}
\label{app:stellar_ev}

\begin{figure}
    \includegraphics[width=0.9\columnwidth]{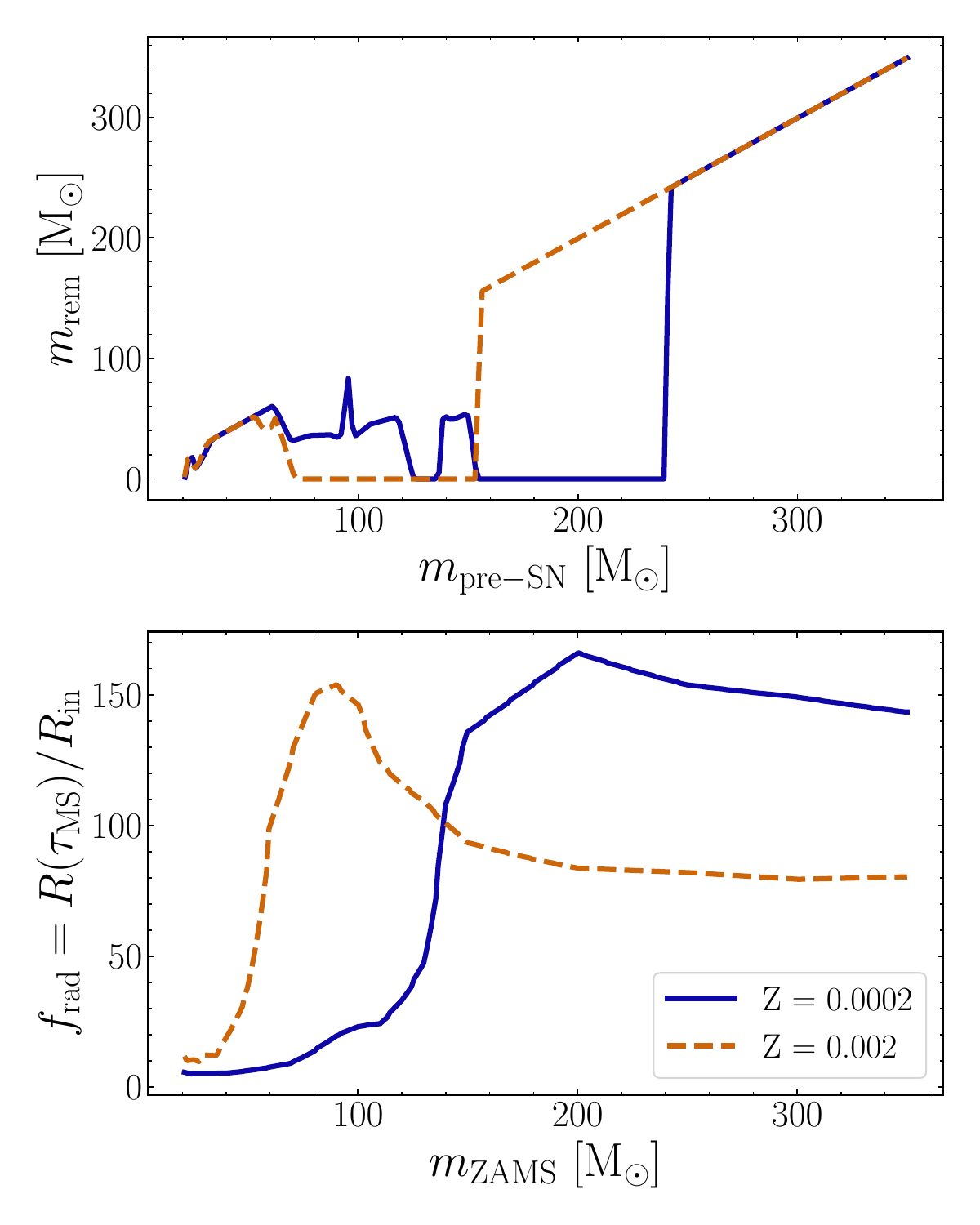}
    \caption{Mass of the compact remnant as a function of the pre-SN stellar mass (upper panel) and $f_{\rm rad}$ as a function of the ZAMS mass (bottom plot) for Z = 0.0002 (blue solid line) and Z = 0.002 (orange dashed line).}
    \label{fig:rem_mass_f_rad}
\end{figure}

\begin{figure}
    \includegraphics[width=\columnwidth]{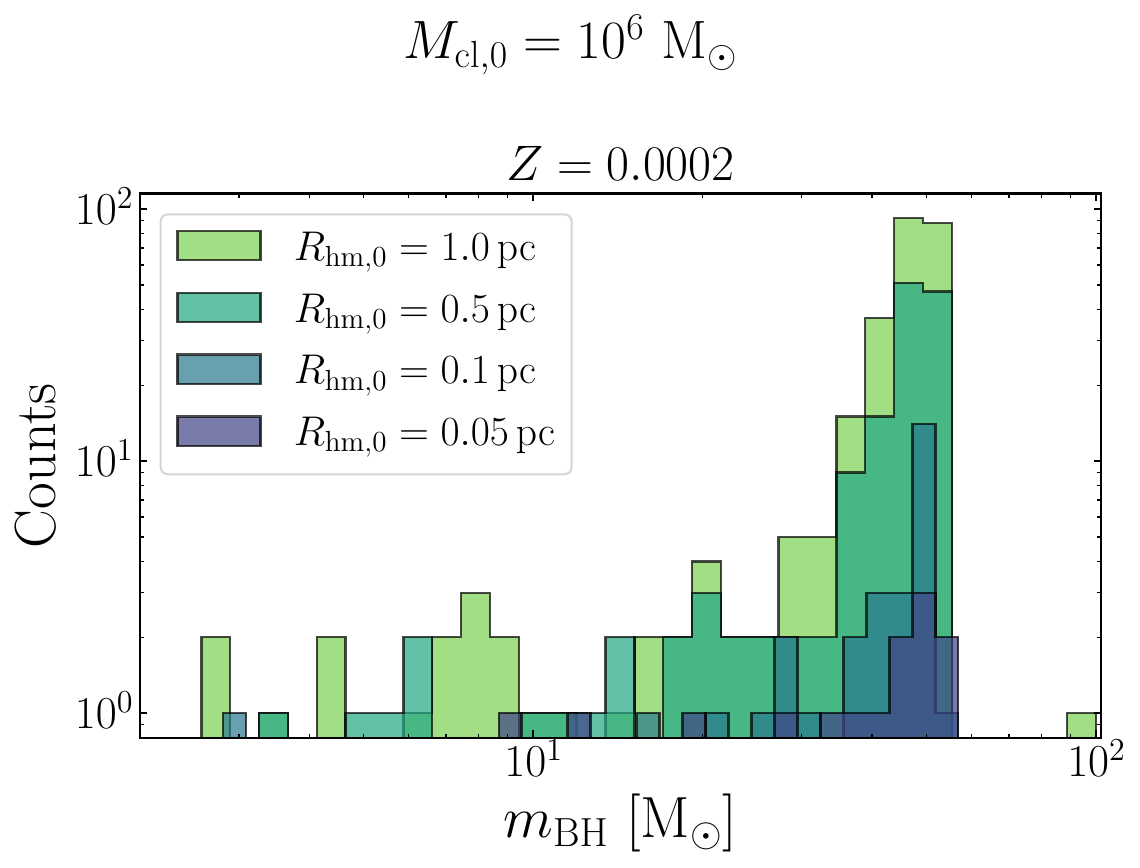}
    \caption{Mass spectrum of stellar-mass black holes for the same clusters considered in Fig. \ref{fig:histo_vms_imbh}. The plot corresponds to metallicity $Z = 0.0002$.}
    \label{fig:histo_stellar_bh}
\end{figure}

As discussed in Sect. \ref{sec:methods}, we extrapolated fitting formulae calibrated on stellar evolution tracks from \sevn-\textsc{Parsec} \citep{Spera2019,Iorio2023,Costa2025} to describe the initial stellar radius ($R_{\rm ZAMS}$; Eq. \ref{eq:r_zams}), the time at which the star begins to expand ($\tau_{\rm MS}$; Eq. \ref{eq:t_evo}) and the size of this expansion ($f_{\rm rad} = R_*(\tau_{\rm MS})/R_{\rm ZAMS}$; lower panel of Fig. \ref{fig:rem_mass_f_rad}), the stellar lifetime ($\tau_{\rm life}$; Eq. \ref{eq:tlife}), and the remnant mass (upper panel of Fig. \ref{fig:rem_mass_f_rad}).

We assumed a constant stellar radius, $R_{\rm ZAMS}$, for $t \leq \tau_{\rm MS}$. After this time, it is increased by a factor $f_{\rm rad}$. As noted when discussing Fig. \ref{fig:vms_mass_growth}, we do not expect a more gradual growth of the stellar radius to significantly affect the final stellar mass, but rather to produce a smoother rise of the mass.

We computed the mass of the compact remnants extrapolating the relation between the pre-SN mass and the remnant mass, i.e, assuming that the remnant mass depends only on the supernova phase (\texttt{rapid} formalism; \citealt{Fryer2012}). This can have an important consequence on the final fate of the VMS. Within the range of initial stellar masses considered for the primary component of the binary (uniform distribution between $100-150 \,\Ms$), a large fraction of stars is expected to fall into the pair-instability regime \citep{Woosley2017,Spera2017}. Under these conditions, the evolution may lead to the formation of low-mass black holes in the case of a PPISN, or to no compact remnant at all in the case of a PISN, despite the high mass of the progenitor stars.

At $Z=0.0002$ ($Z = 0.002$), the fraction of stars entering the PISN regime increases from 1\% (4\%) to 38\% (63\%) as the half-mass radius varies from $R_{\rm hm,0}$ = 0.05 pc to $R_{\rm hm,0}$ = 1 pc. The fraction of stars producing stellar-mass black holes varies, instead, from $\sim\,3$\% to 54\% at $Z=0.0002$ for $R_{\rm hm,0}$ = 0.05 pc and $R_{\rm hm,0}$ = 1 pc respectively. For the case $Z=0.002$, no stellar-mass BH are formed. All these values refer to clusters with an initial mass $M_{\rm cl,0} = 10^6 \,\Ms$ (Figs. \ref{fig:vms_mass_growth}; \ref{fig:histo_vms_imbh};\ref{fig:histo_stellar_bh}). 

This behavior results from the combined effect of our extrapolation of the \sevn-\textsc{Parsec} stellar tracks, initial conditions, and the mass loss induced by stellar winds and stellar collisions that allows stars to enter or exit the PISN regime depending on the number of interactions and their lifetime.

These values should not be interpreted as absolute rates, as they depend on the adopted initial conditions.

\FloatBarrier 

\section{Impact of mass loss}
\label{app:mass_loss}

\begin{figure*}[!ht] 
    \centering
    \includegraphics[width=0.9\textwidth]{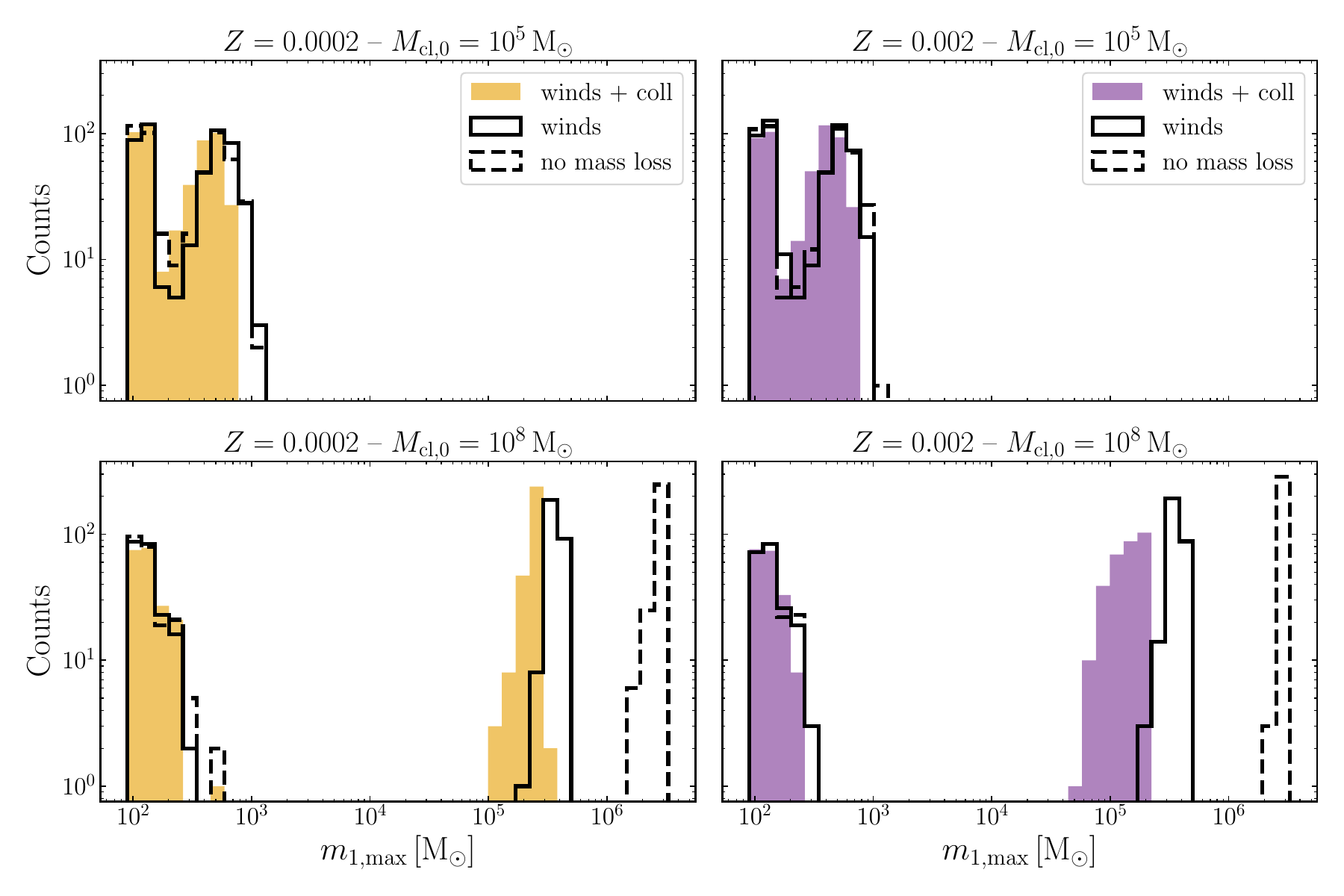}
    \caption{Distribution of the maximum stellar mass for clusters with mass $M_{\rm cl,0} = 10^5 \,\Ms$ (upper panels) and $M_{\rm cl,0} =10^8\,\Ms$ (lower panels), and half-mass radius $R_{\rm hm,0} = 0.5$ pc, for metallicities $Z = 0.0002$ (orange) and $Z = 0.002$ (violet). The filled histograms include mass loss driven by both stellar winds and collisions. The black solid lines account only for stellar-wind mass loss, while the black dashed lines correspond to the case in which no mass loss is considered. }
    \label{fig:histo_winds}
\end{figure*}

In our work, we considered mass loss driven by both stellar winds and collision.

To model mass loss via stellar winds, we adopted the prescription by \cite{Vink2018}:

\begin{equation}
\begin{aligned}
\log\left(\frac{\dot{m_*}}{\,\Ms\,{\rm yr}^{-1}}\right) &= -9.13 \\
&\quad + 2.1\log\left(\frac{m_*}{\,\Ms}\right) \\
&\quad + 0.74\log\left(\frac{Z}{{\rm Z}_{\odot}}\right).
\end{aligned}
\end{equation}

The mass loss in stellar collision and mergers is parametrized as in \cite{Glebbeek2013}:
\begin{equation}
f_{\rm loss} = D\,\frac{q}{(q+1)^2},
\end{equation}
with $q = m_2/m_1 < 1$ and $D = 0.3$.

In addition, we ran some specific simulations to assess the effect of our assumptions.

Figure \ref{fig:histo_winds} shows the distribution of the maximum primary mass $m_{1,{\rm max}}$ obtained in clusters with $M_{\rm cl,0} = 10^5,10^8 \,\Ms$ for different metallicities ($Z = 0.0002,0.002$) at $R_{\rm hm,0} = 0.5$ pc.

Three different prescriptions for  mass loss are compared: stellar wind and collision-driven mass loss, mass loss via stellar winds only, no mass loss.

The comparison highlights the strong impact of mass loss on the maximum stellar masses reached in the simulations. When no mass loss is included, the runaway growth driven by stellar collisions can produce extremely massive stars, reaching $> 10^{6}\mathrm{M_\odot}$ in the most massive clusters (lower panels). Including stellar winds only already reduces the high-mass tail of the distribution, limiting the growth of the most massive objects. When both winds and collisional mass loss are taken into account, the maximum masses are further suppressed, shifting the distributions to significantly lower values ($\lesssim 10^5 \,\Ms$).

This effect becomes particularly evident in the most massive clusters ($M_{\rm cl,0}=10^{8}\,\Ms$), where runaway growth would otherwise lead to the formation of extremely massive stars in the absence of mass loss. The combined effect of winds and collision-induced mass loss efficiently regulates the growth of the most massive objects, preventing the formation of the most extreme masses predicted in the no-mass-loss scenario. The impact of stellar winds is metallicity dependent, with higher metallicity models ($Z=0.002$) showing a stronger suppression of the high-mass tail due to the increased efficiency of wind-driven mass loss.

\FloatBarrier 
\clearpage

\end{appendix}

\end{document}